\documentclass[apj]{emulateapj}
\slugcomment{{\sc Accepted to AJ:} April 5, 2011} 

\begin{document}

\title{\textit{Swift} UV/Optical Telescope Imaging of Star Forming
  Regions in M81 and Holmberg IX}

\author{E.~A. Hoversten\altaffilmark{1}, C. Gronwall\altaffilmark{1}, D.~E. Vanden Berk\altaffilmark{2}, A.~R. Basu-Zych\altaffilmark{3}, A.~A. Breeveld\altaffilmark{4}, P.~J. Brown\altaffilmark{5}, N.~P.~M. Kuin\altaffilmark{4}, M.~J. Page\altaffilmark{4}, P.~W.~A. Roming\altaffilmark{6}, and M.~H. Siegel\altaffilmark{1}} 

\altaffiltext{1}{Department of Astronomy \& Astrophysics, The Pennsylvania State
  University, 525 Davey Laboratory, University Park, PA 16802}
\altaffiltext{2}{Physics Department, St. Vincent College, Latrobe, PA 15650}
\altaffiltext{3}{NASA/Goddard Space Flight Center, Greenbelt, MD 20771}
\altaffiltext{4}{Mullard Space Science Laboratory/UCL, Holbury St. Mary, Dorking, Surrey RH5 6NT}
\altaffiltext{5}{Department of Physics \& Astronomy, University of Utah, Salt Lake City, UT 84112}
\altaffiltext{6}{Space Science \& Engineering Division, Southwest Research Institute, 6220 Culebra Road, San Antonio, TX 78238}

\shorttitle{\textit{SWIFT} UVOT OBSERVATIONS OF M81 \& HOLMBERG IX}
\shortauthors{HOVERSTEN ET AL.}

\begin{abstract}
We present \textit{Swift} UV/Optical Telescope (UVOT) imaging of the galaxies M81 and Holmberg IX.  We combine UVOT imaging in three near ultraviolet (NUV) filters (uvw2: 1928 \AA, uvm2: 2246 \AA, and uvw1: 2600 \AA) with ground based optical imaging from the Sloan Digital Sky Survey to constrain the stellar populations of both galaxies.   Our analysis consists of three different methods.  First we use the NUV imaging to identify UV star forming knots and then perform SED modeling on the UV/optical photometry of these sources.  Second, we measure surface brightness profiles of the disk of M81 in the NUV and optical.  Last we use SED fitting of individual pixels to map the properties of the two galaxies.  In agreement with earlier studies we find evidence for a burst in star formation in both galaxies starting $\sim200$ Myr ago coincident with the suggested time of an M81-M82 interaction.  In line with theories of its origin as a tidal dwarf we find that the luminosity weighted age of Holmberg IX is a few hundred million years.  Both galaxies are best fit by a Milky Way dust extinction law with a prominent 2175 \AA\ bump.  In addition, we describe a stacked median filter technique for modeling the diffuse background light within a galaxy, and a Markov chain method for cleaning segment maps generated by SExtractor.
\end{abstract}


\section{Introduction}

Discovered by Johann Bode in 1774 and appearing in Messier's original catalog \citep{Messier81}, M81 (NGC 3031, \textit{Bode's Galaxy}) has been attracting the attention of astronomers for centuries.  It is among the nearest grand design spiral galaxies and its large angular extent ($14' \times 27'$) lends itself to in depth study as well as majestic images.

M81 is classified as an SA(s)ab galaxy \citep{deV91} meaning that it is a spiral galaxy with smooth, tightly wound arms and a prominent bulge and without a bar or ring structure.  However \citet{Elmegreen95} show evidence for a small bar.  M81 also possesses an active nucleus which is classified as a LINER \citep{Heckman80} which has been monitored by \textit{Swift} \citep{Pian10}.  The Cepheid distance to M81 is $3.63\pm0.33$ Mpc \citep{Freedman01}.  \citet{Gordon04} estimates the current star formation rate (SFR) over the whole galaxy as 0.31, 0.38, and 0.89 $M_\odot$ yr$^{-1}$ based on the integrated UV, H$\alpha$, and infrared luminosities (these values differ due to the presence of dust obscured star formation). This is slightly lower than recent estimates of the SFR in the Milky Way which range from around 1 \citep{Robitaille10} to 4 \citep{Diehl06} $M_\odot$ yr$^{-1}$.  The $V$ band extinction has been measured to be $A_V \sim 0.5$-1 with some spatial variation \citep{Kaufman87,Garnett87,Kong00}.

Observations of M81 have been made across the electromagnetic spectrum from the X-ray to the radio.  This includes a number of multi-wavelength campaigns.  One area of particular interest is the cross calibration of SFR indicators from UV, H$\alpha$, and infrared measurements \citep{Devereux95,Gordon04,PG06}.  Using these techniques \citet{Gordon04} found evidence for obscured star formation in M81.  Similarly \citet{PG06} found that 50\% of the total star formation in M81 is obscured by dust, ranging from 60\% in the bulge to 30\% in the outer regions suggesting a radial dust gradient.  \textit{Herschel} observations from 70-500 $\mu$m \citep{Bendo10} show that dust emission in the infrared traces the spiral arms of M81.  However these results also suggest that $\sim 30$\% of the dust emission at 70 $\mu$m comes from dust heated by evolved stars rather than star formation indicating that infrared SFR indicators may need to be revised.

M81 is the largest spiral galaxy in the M81 Group of galaxies.  \citet{Karachentsev05} counts 29 (plus 6 possible) members of the M81 Group, including M82, NGC 3077, and Holmberg IX.  Radio imaging by \citet{Yun94} revealed filamentary \ion{H}{1} structures connecting M81, M82, and NGC 3077 indicating recent tidal interactions.  Based on numerical simulations \citet{Yun99} estimate that the closest approach to M81 occurred 220 Myr ago for M82 and 280 Myr for NGC 3077.  \citet{Boyce01} suggested that at least three members of the M81 Group (Holmberg IX, Arp's Loop, and the Garland) may be tidal dwarf galaxies, formed as a result of interactions between M81, M82, and NGC 3077.  By contrast \citet{Hammer07} argue that the Milky Way has not experienced any significant merger event for the past $\sim10$ Gyr, which is not the case for most spiral galaxies.  

The idea that new galaxies, known as tidal dwarfs, could be formed from tidal debris from interacting galaxies was first proposed by \citet{Zwicky56}, and observational evidence has been accumulating for their existence \citep[e.g.][]{Sabbi08}.  In addition to \citet{Yun99} there are many other examples of numerical N-body simulations of interacting galaxies which show evidence for the formation of tidal dwarf galaxies.  Modeling of the interacting system Arp 245 by \citet{Duc00} suggests that a star-forming optical condensation, A245N, may be the progenitor of a future tidal dwarf.  In more general simulations by \citet{BDM03} it is shown that tidal dwarf galaxies can not only form in galaxy interactions, but a third of them remain on nearly circular orbits allowing them to avoid or substantially delay reabsorption by the dominant galaxy.  This suggests that a significant fraction of satellite dwarf galaxies could be of tidal origin.  As another example, similar techniques have been applied to address the topic of dark galaxies.  The proposed dark galaxies \citep[e.g.][]{CF88} are characterized by high dynamical mass, as indicated by \ion{H}{1} emission, but lack an optical counterpart and are thus presumably dark matter dominated.  \citet{DB08} applied similar N-body modeling to the candidate dark galaxy VIRGOHI 21 to show that these dark galaxies may be more simply explained by tidal origins.

The tidal streams in the M81 Group have prompted a number of investigations into stellar populations residing in the stream.  New young stellar systems have recently been discovered in the M81 debris field using both UV \citep{deMello08} and optical \citep{Davidge08, Mouhcine09,Barker09} imaging.    Unfortunately the field of view of the UV imaging in this paper does not extend across the full debris field and this avenue of inquiry cannot be pursued here.

However the candidate dwarf galaxy Holmberg IX does fall within the field for our M81 imaging.  Despite its location just 10' east of the nucleus of M81 the galaxy Holmberg IX was not cataloged until 1959 \citep[DDO 66,][]{VDB59}.  It received its current designation and was briefly described by \citet{Holmberg69} and was further described by \citet{Bertola74} who drew comparisons to the Small Magellanic Cloud in size but noted that its surface brightness is four magnitudes lower.

\citet{Sabbi08} make a strong argument that Holmberg IX is a tidal
dwarf galaxy.  Using deep \textit{Hubble Space Telescope} (HST) imaging they
constructed color-magnitude diagrams of the resolved stellar
population in Holmberg IX.  They found that Holmberg IX is dominated
by stars with ages $\lesssim 200$ Myr, consistent with the proposed
time of the M81-M82 interaction.  There is also a population with ages
greater than 1 Gyr which they argue is more likely associated with
the halo of M81 or other tidal debris.  The combination of its
location in one of the main tidal \ion{H}{1} streams near M81 and that
it possesses the youngest mean stellar population of any nearby galaxy
lead them to conclude that Holmberg IX is of recent tidal origin.  Similar analysis of resolved stellar populations by \citet{Weisz08} is also indicative of a tidal dwarf origin for Holmberg IX.

This paper presents near UV imaging of the field of M81 and Holmberg IX from the \textit{Swift} Ultraviolet/Optical Telescope \citep[UVOT;][]{UVOT}.  We combine the UVOT data with optical imaging to investigate the UV properties of star formation in the two galaxies.  The organization of this paper is as follows: In Section 2 observations and data reduction are described, Section 3 provides a description of the stellar population modeling, Section 4 describes data analysis, and Section 5 includes a discussion of the results.

\begin{deluxetable*}{lccccccccc}
\tabletypesize{\tiny}
\tablecaption{\textit{Swift} UVOT observations of M81}
\tablewidth{0pt}
\tablehead{
 & & & \multicolumn{6}{c}{Exposure Time (s)} & \\
\colhead{Name} & \colhead{ObsID} & \colhead{Date} & \colhead{uvw2} & \colhead{uvm2} & 
\colhead{uvw1} & \colhead{$u$} & \colhead{$b$} & \colhead{$v$} & \colhead{M81 offset}\\
 & & & \colhead{1928\ \AA} & \colhead{2246\ \AA} & \colhead{2600\ \AA} & 
\colhead{3465\ \AA} & \colhead{4392\ \AA} & \colhead{5468\ \AA} & \colhead{arcmin}\\
 & & & \colhead{$2\farcs 92$} & \colhead{$2\farcs 45$} & 
\colhead{$2\farcs 37$} & \colhead{$2\farcs 37$} & \colhead{$2\farcs 19$} & \colhead{$2\farcs 18$} & }
\startdata
M81 & 00035059001 & 2005-04-21 &    0 &    0 &    0 &  102 &   45 &  123 & 1.13\\
M81 & 00035059002 & 2005-08-25 & 1596 & 1172 &  778 &  390 &  334 &  392 & 1.99\\
M81 & 00035059003 & 2006-06-23 & 1469 &  930 &  731 &  367 &  367 &  367 & 0.97\\
M81X9 & 00035335001 & 2006-07-07 & 3953 & 2591 & 1976 &  987 &  987 &  987 & 12.58\\
M81X9 & 00035335002 & 2006-07-08 & 1077 &  665 &  539 &  267 &  267 &  267 & 11.94\\
\enddata
\label{tab:uvotobs}
\tablecomments{Columns are the name of the observing target, the \textit{Swift} observation ID number,
  the start date of the observation, the exposure times in each
  filter for each observation, and the offset of the pointing from the center of M81 in arcminutes.  The central wavelengths (assuming a flat spectrum) and the PSF FWHMs from \citet{UVOTII} are given for each filter.  This table is available in its entirety in a machine-readable form in the online journal.  A portion is shown here for guidance regarding its form and content.} 
\end{deluxetable*}

\section{Observations and Reductions\label{sec:obs}}

Observations of M81 and Holmberg IX were made with UVOT.  UVOT is one of
three telescopes onboard the \textit{Swift} spacecraft \citep{Swift}.  UVOT is a 30 cm telescope with two grisms and seven broadband filters.  The
central wavelengths and PSFs of the filters are given in Table
\ref{tab:uvotobs}.  For a more detailed discussion of the filters, as
well as plots of the responses, see \citet{UVOTcal} as well as the updates in \citet{Breeveld10b}.

UVOT observations of M81 and Holmberg IX were made between 2005 April 21 and 2009 September 4.  Observations of Holmberg IX and
the eastern edge of M81 are deepest due to various
\textit{Swift} programs to monitor the ultra-luminous X-ray source M81
X-9 \citep[see][]{LaParola01} which is located in
Holmberg IX around 12.5 arcminutes east-northeast of the center of
M81.  Some observations were taken in an unbinned mode with a plate
scale of 0.5 arcseconds per pixel.  However the majority were observed
in a $2\times 2$ binning mode with a scale of 1 arcsecond per pixel to
reduce the amount of telemetry from the spacecraft.  Not all available
observations have been used.  Some observations were omitted either because the UVOT pipeline did not find an aspect solution, or the observations were taken in a mode where UVOT was read out at a non-standard frame time.  The UVOT software does not allow the combinations of observations with different frame times, as this greatly complicates analysis of the images.  These problematic observations have been discarded.  A list of observations used in this paper is shown in Table \ref{tab:uvotobs}.

Reductions of the UVOT data largely follow the methods described in
\citet{Hoversten09}.  UVOT data processing is described in the UVOT
Software
Guide\footnote{\texttt{http://heasarc.gsfc.nasa.bov/docs/swift/analysis}}.
However much of the early observations were processed with a version
of the UVOT pipeline in which exposure maps are not aspect corrected.
To address this problem new exposure maps were created from the image
files, UVOT housekeeping data, and spacecraft auxiliary files (all
available in the archive) using the current version of
\texttt{UVOTEXPMAP} from the publicly available UVOT FTOOLS (HEAsoft 6.6.1)
\footnote{\texttt{http://heasarc.gsfc.nasa.gov/docs/software/lheasoft/}}.
Image files and exposure maps were summed using \texttt{UVOTIMSUM}
(also a UVOT FTOOL).  This involves two flux conserving interpolations
of the images.  The first of these converts from the raw frame to sky
coordinates and the second occurs when summing the images.  As a
result of this interpolation the final summed images are on the scale
of the most coarsely binned image, which is $2\times 2$ for these
observations.  A correction is applied in the processing for known bad
pixels.  

Cosmic ray corrections are not necessary for UVOT images.  Individual events are identified and centroided upon in each UVOT frame and placed into an image at a later stage.  A cosmic ray hitting the detector will register one or a few counts after centroiding, rather than the thousands of counts which occur in CCDs operating in the usual integrating modes.  As a result, cosmic rays are simply a part of the background in UVOT images.

UVOT is a photon counting instrument and as such is subject to
coincidence loss, which is also called ``pile-up'' in X-ray astronomy.
Coincidence loss occurs when two or more photons arrive at nearly the
same location within the same CCD frame \citep{Fordham00}.  Multiple
incident photons can be mistaken for a single photon resulting in a
systematic undercounting of the flux.  As coincidence loss is a
function of brightness it affects the linearity of the detector. For
point sources a coincidence loss correction for UVOT has been determined by \citet{UVOTcal} with the photometric errors due to coincidence loss explored by \citet{Kuin08}.  However for extended sources no such correction yet exists.

\begin{figure}
\plotone{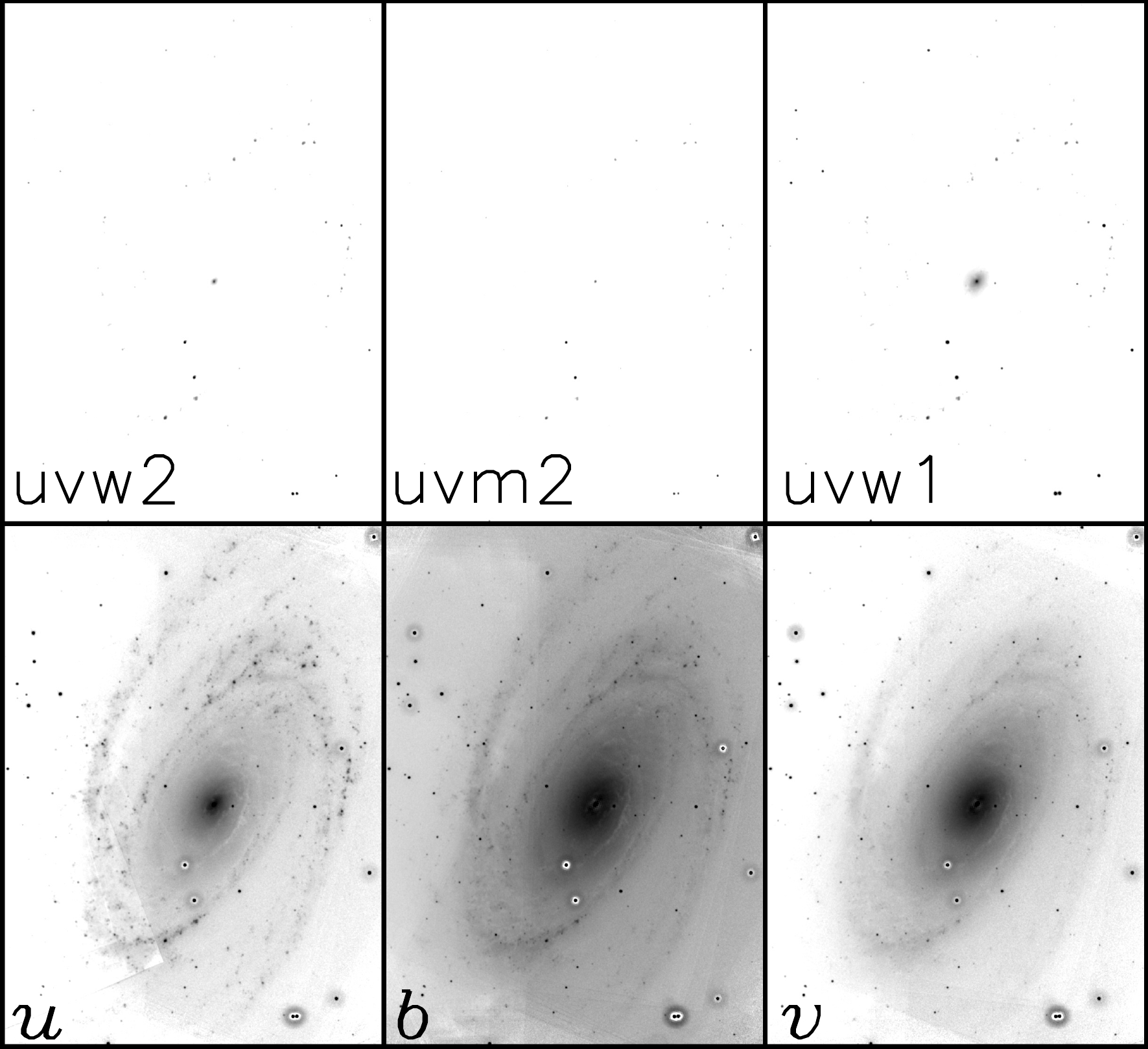}
\caption{UVOT count rate images of M81 by filter.  All images are
  scaled identically with a logarithmic stretch.  Non-white areas have a count rate higher than
  0.028 cts s$^{-1}$ pixel$^{-1}$ which indicates that coincidence
  loss effects are larger than 1\% (for images with $2 \times 2$ binning as discussed in the text).  For the uvw2, uvm2, and uvw1
  filters only the brightest star forming regions are affected, in
  addition to a few stars and the inner regions of the bulge.  However
  in the $u$, $b$, and $v$ filters coincidence loss is a problem
  across the whole disk of the galaxy. The white rings around bright
  sources are an artifact of the UVOT onboard centroiding algorithm in the
  high count rate regime.}
\label{fig:coitile}
\end{figure}

It is therefore important to quantify the effects of coincidence loss
in the M81 images.  Figure \ref{fig:coitile} shows the summed images in each
filter in observed counts per second with the same scaling across all
images.  Counts per frame is the unit which is directly relevant for
coincidence loss.  However all of the M81 images were taken with a frame time of 11 ms so there is a simple conversion to counts per second.  Figure \ref{fig:coitile} shows that, in general, count
rates across the images in $u$, $b$, and $v$ are much higher than for
the NUV filters.  \citet{UVOTII} modeled the UVOT response numerically for point
sources with high backgrounds.  They showed that for background count
rates less than 0.007 cts s$^{-1}$ pix$^{-1}$ the flux of extended
sources is off by at most 1\% due to coincidence loss.  Because our
images of M81 are binned $2\times 2$ the relevant count rate is four times
larger.  In any part of an image where pixels have count rates in
excess of 0.028 cts s$^{-1}$ pix$^{-1}$ the effects of
coincidence loss need to be either explicitly calculated if possible
or disregarded. 

Figure \ref{fig:coitile} reveals that coincidence loss is only a problem at
the core of M81 and in the brightest star formation regions and
foreground stars in the NUV filters.  The core of M81 is therefore
excluded in our analysis, and bright star forming regions are
corrected using the point source methods described in \citet{UVOTcal}
which is further discussed in \S \ref{sec:phot}.  However the problem
is much more serious in the $u$, $b$, and $v$ filters where the per
pixel count rates are high across the disk of M81.  For these images a
coincidence loss correction is untenable at this point in time.

To avoid these severe coincidence loss issues ground based imaging was used in place of the UVOT $u$, $b$, and $v$
images.  Optical images were obtained from the Sloan Digital Sky Survey
\citep[SDSS;][]{York00}.  Imaging was done
in the $ugriz$ filter system described by \citet{Fukugita96}.  SDSS
imaging is performed using CCDs in the more common integrating mode
which does not suffer from coincidence loss.  Imaging of M81 was done
on November 20 and 30, 2003 in runs 4264 and 4294.  Corrected imaging
frames (\texttt{fpC} files) and frame summary (\texttt{tsField}) files
were retrieved from the SDSS Data Archive Server.

SDSS imaging is done in drift scanning mode with an exposure time of
54 seconds.  Individual corrected frames were bias subtracted and sky
subtracted.  The sky values were estimated using the median value of
the frame.  For frames overlapping the central regions of M81 the
background level of adjacent frames from the same run and camera
column were used instead.  The airmass and extinction coefficients are
given in the \texttt{tsField} file and included in the flux determination.
Individual frames were rebinned to match the 1'' pixel scale of the
UVOT images (the SDSS pixels are 0.4''), rotated, and registered using
publicly available IDL routines \citep{Landsman93}.  Individual frames
were then combined into single images for each filter, and flux calibrated using information from the relevant \texttt{tsField} files.

The first step in the flux calibration of the UVOT images is to divide the stacked UVOT images by the exposure maps to create count rate images.  Flux calibration of the UVOT images is described later, depending on the type of analysis being performed.

\begin{figure*}
\plotone{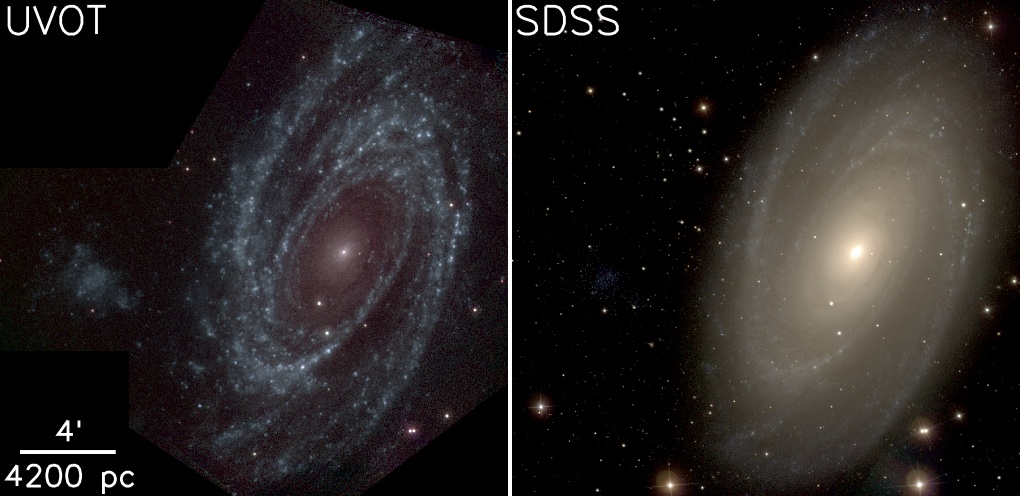}
\caption{False color images of M81.  At left is an ultraviolet image
  of M81 constructed from \textit{Swift} UVOT images using the uvw1 (red), uvm2
  (green), and uvw2 (blue) filters.  At right is the optical image
  from SDSS imaging using the $i$ (red), $r$ (green), and $g$ (blue)
  images.  At the lower left the scale of the image is given in both arcminutes and the corresponding physical scale in parsecs assuming a distance of 3.63 Mpc to M81.  In both images the scaling is logarithmic.}
\label{fig:uvopt}
\end{figure*}

\begin{figure}
\plotone{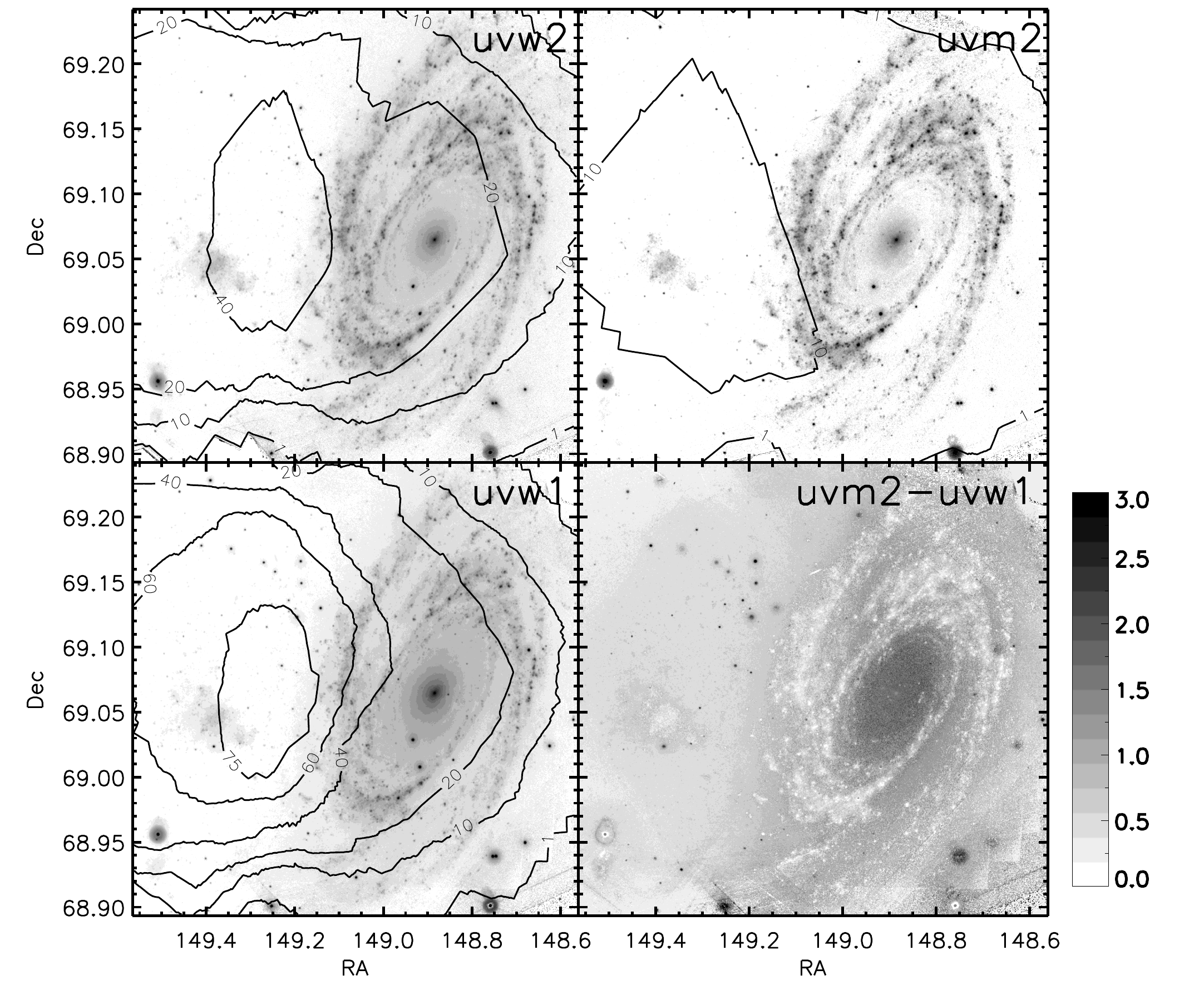}
\caption{Exposure maps for UVOT observations of M81 in the uvw2 (top
  left), uvm2 (top right), and uvw1 (bottom left) filters. The panels
  show contours in exposure time as labeled in kiloseconds overlaid on the respective UVOT images with logarithmic scaling. Observations are deepest near Holmberg IX due to extensive \textit{Swift} monitoring of M81 X-9.  At bottom right the uvm2-uvw1 color of the field is shown, with the color bar at right indicating the color associated with the levels of shading.  Exposure time in the SDSS images is 54 seconds, except in the narrow regions where stripes overlap.}
\label{fig:expmap}
\end{figure}

False color images constructed from the summed UVOT and SDSS images
are shown in Figure \ref{fig:uvopt}.  In the left panel the UV image
reveals pronounced spiral arms and relatively little emission in the
central regions.  As discussed later, the uvw1 filter has an extended red tail, so the red component at the center of M81 in the UV image may be dominated by optical light.  In the optical image the arms are still present, but there is much more diffuse emission throughout.  Holmberg IX stands out in the UV image, but is difficult to see in the optical.  UVOT exposure maps
for the NUV filters are shown in Figure \ref{fig:expmap}.
Observations are deeper in uvw2 and uvw1 and are deepest near Holmberg
IX due to multiple \textit{Swift} observing campaigns of the X-ray
source M81 X-9.

\subsection{Diffuse Background Removal\label{sec:dback}}
 
One of the goals of this paper is to find and measure the star forming regions
in M81.  The degree of success to which this goal is achieved is
related to how well the diffuse galaxy background can be subtracted.
There are a number of approaches to modeling the diffuse background.
One is to use a median filter where the value of each pixel is
replaced by the median pixel value of a circular or rectangular window
centered on the pixel.  Another is the ring median filter
\citep{Secker95}, which removes objects with a scale length smaller
than the inner radius of the ring.

\begin{figure*}
\plotone{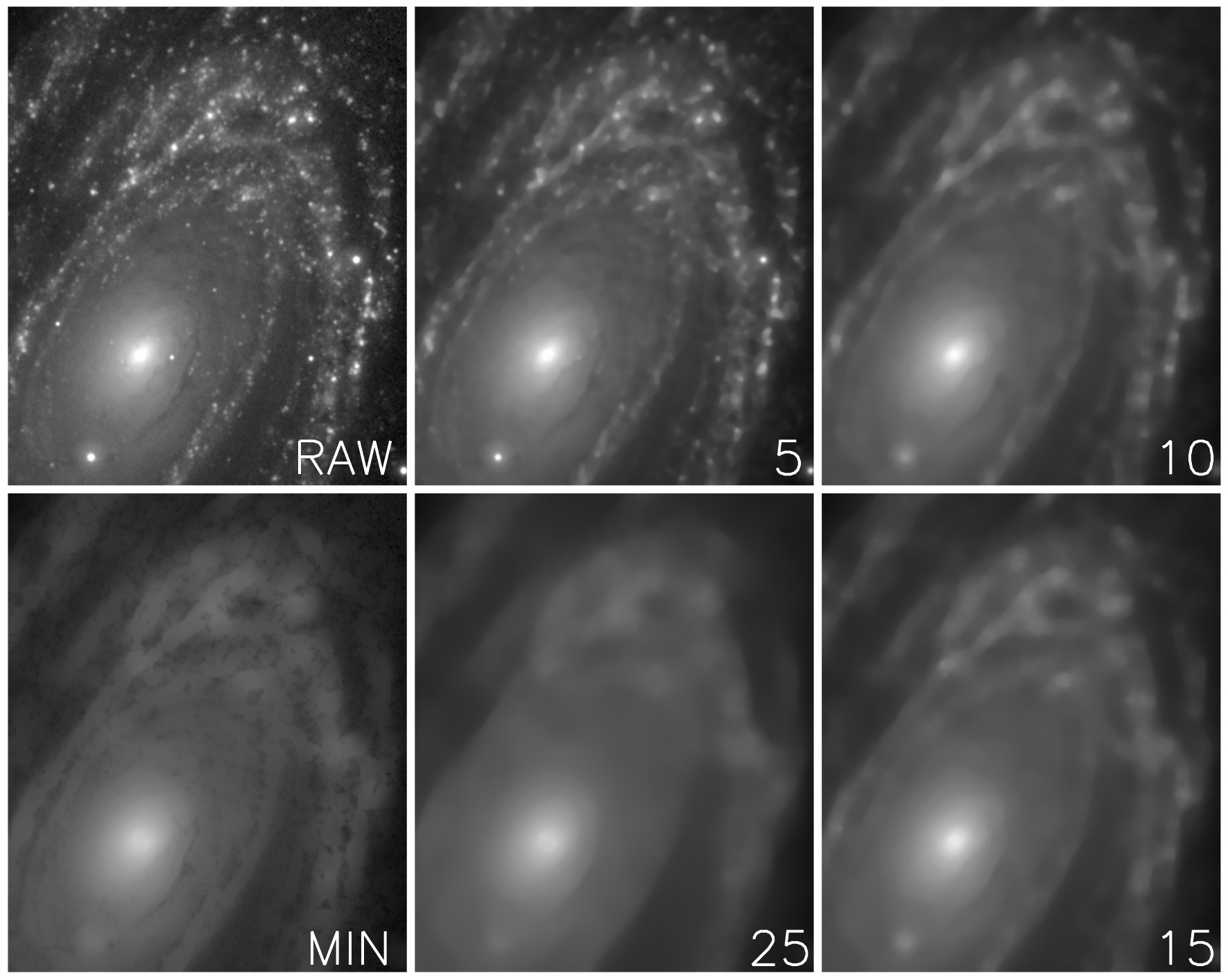}
\caption{Circle median filtered uvw1 images of a portion of M81.
  Clockwise from upper left: unfiltered image, images with filter
  radii of 5, 10, 15, and 25 pixels, and combined minimum filter
  described in the text. The logarithmic scaling is identical in all panels.  The 5 pixel image fails to remove all of the point sources while narrow
  dust lanes are washed out in the 25 pixel image.  The combined
  minimum image eliminates point sources while retaining small scale
  structure of the diffuse background.}
\label{fig:filter_tile}
\end{figure*}

Median filtering has a number of drawbacks.  For one, the scale of
structure in M81 varies from region to region.  There are narrow dust
lanes as well as broad spiral arms.  If the size of the filter window
is too small it will fail to remove the star forming regions from the
background image and these regions of interest will subsequently be
subtracted out.  If the filter window is too large the fine structure
of the spiral arms will be washed out in the background image.  This
leads to large regions in the background subtracted image with large
negative fluxes where small star forming regions may fail to be
identified.  These issues are demonstrated in Figure
\ref{fig:filter_tile} for circle median filtering of the uvw1 image of
M81.  Circle median filtering, as shown in Figure \ref{fig:filter_tile}, is more computationally expensive than the ring median filtering described in \citep{Secker95}.  However ring median filtering has the drawback that background images have a visibly noticeable noise pattern in them.  For this reason we use circle median filtering throughout this paper.

The solution is to use multiple circular median filtered images of
varying filter radii to estimate the diffuse background of M81.  In
this way regions with different scales of structure can be handled
appropriately without sacrificing accuracy in another region of the
galaxy.  To do this the \texttt{rmedian} task was used in IRAF/PyRAF
to make circle median filtered images for each UVOT filter with an inner radius of 0 and outer radii of 5, 6, 7, 8, 10, 13, 15, 20, and 25 pixels.  At a distance
of 3.63 Mpc \citep{Freedman01} the 1'' pixels correspond to a physical
distance of 17.6 pc.  \citet{Youngblood99} measure the size
distribution of \ion{H}{2} regions in irregular galaxies and find that
in some galaxies \ion{H}{2} regions can have diameters as large as
500 pc.  Along with \citet{Strobel91} they find a weak correlation
between the characteristic size of \ion{H}{2} regions and galaxy
luminosity in irregulars with \ion{H}{2} regions increasing in size
with increasing galaxy luminosity.  Applying this relationship to M81,
although it is not an irregular, suggests that the characteristic diameter
of \ion{H}{2} regions should be around 150 pc.  As we expect the UV star
formation regions to be one and the same as \ion{H}{2} regions we use
this information to guide the upper limit on the size of the median
filter radius.  

With the higher resolution of {\it HST} \citet{Pleuss00} and \citet{Scoville01} found that the sizes of \ion{H}{2} regions are generally much smaller than 150 pc.  In particular \citet{Pleuss00} investigated the effects of the resolution of observations on the measured sizes of \ion{H}{2} regions.  They found that the measured diameters increased as the resolution of the imaging decreased.  A 25 pixel radius corresponds to 440 pc at the distance of M81.  While HST has revealed that the physical sizes of \ion{H}{2} regions are much smaller, this is a sensible upper limit for the radius of median filters for UVOT imaging where the regions will be unresolved or blended.

These circular filter images,
along with the original unfiltered image, are read into a data
cube. For each image pixel the diffuse background is equal to the
minimum value of the corresponding pixels in the data cube.  This can be
understood by considering two extreme cases.  First, consider a pixel
at the center of an isolated point source.  The raw image pixel will
have a large number of counts, but the same pixel in the median filter
image with the largest circular radius will have a value close to the
sky background.  On the other extreme is a pixel in a heavily
extincted dust lane between spiral arms in the M81 image.  The raw
image pixel will have a low value due to the large extinction.  The
median filtered image pixels will have larger values because they take
the median over adjacent, brighter pixels.  In either case the minimum
value of a pixel is the most correct model of the diffuse background.

To improve the quality of the diffuse background image while
minimizing computational cost, for each pixel the column through the
data cube is interpolated quadratically at 1 pixel intervals in circle median filter radius.  The minimum value of this interpolation is used in the diffuse background image.  The radii of circular median images listed above were selected to ensure high enough sampling that the interpolation was smoothly varying. A byproduct of this procedure is a mapping of the scale of structure in the galaxy based on the size of the interpolated circular median radius that has the smallest value.  The combined minimum median filtered image of M81 in the UVW1 band is shown in the lower left panel of Figure \ref{fig:filter_tile}.  A comparison with the median filtered images of various radii in Figure \ref{fig:filter_tile} demonstrates the success of this technique in removing point sources while simultaneously preserving more of the structure of the diffuse background.

\begin{figure*}
\plotone{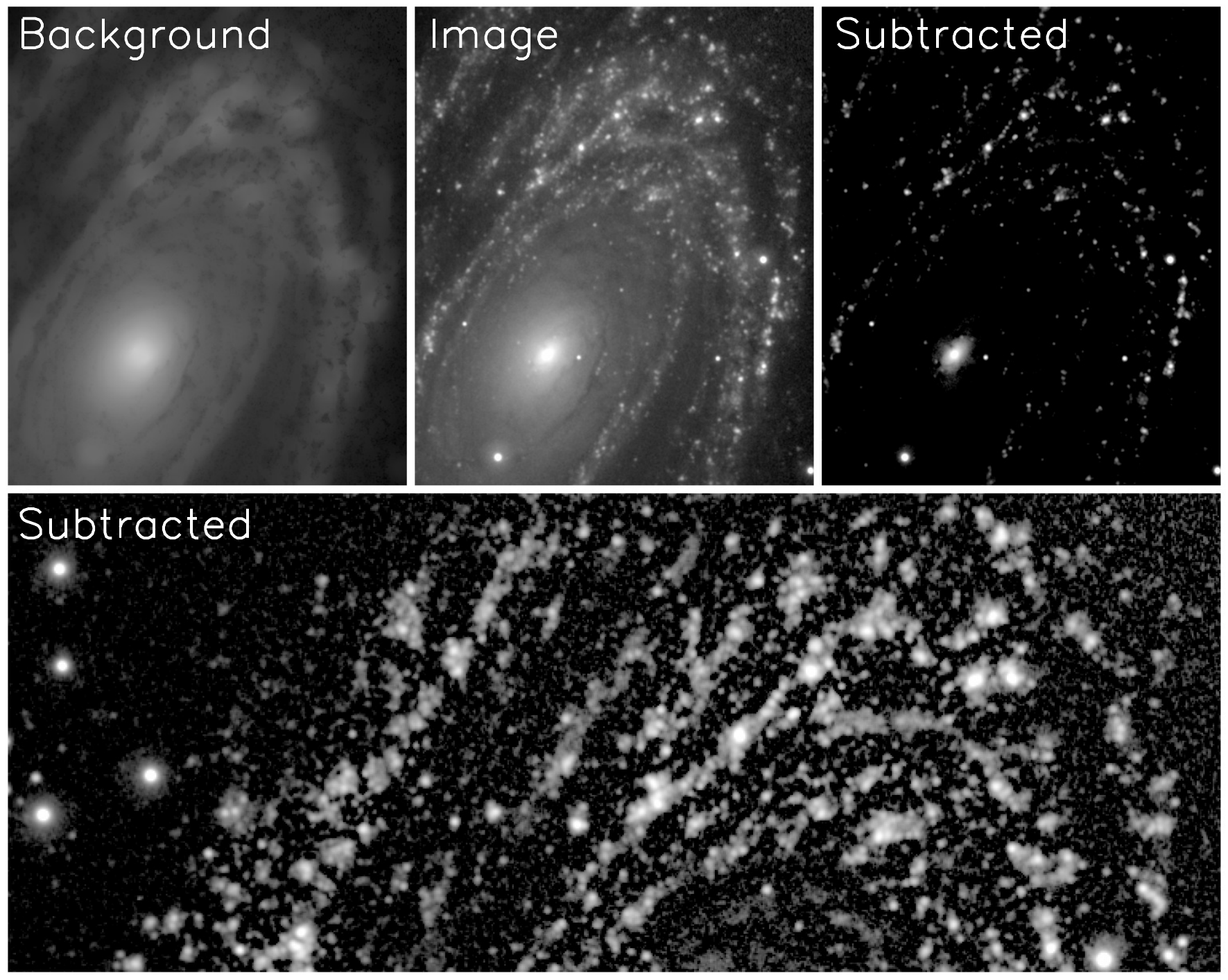}
\caption{An example of the performance of the diffuse background subtraction in the uvw1 filter.  At top left is the portion of the diffuse background image from Figure \ref{fig:filter_tile}.  At top center is the raw count rate image, and at top right is the diffuse background subtracted count rate image.  The three panels have been logarithmically scaled identically.  The bottom panel is a zoom of the diffuse background subtracted count rate image.  It has been rescaled to bring out the noise level of the image.}
\label{fig:back_subt}
\end{figure*}

An image of the star forming regions of M81 can then be constructed by
subtracting the diffuse background image from the original image in
each filter.  These images are effectively sky subtracted due to the
manner in which the background was measured.  Figure \ref{fig:back_subt} shows the diffuse background subtraction in the uvw1 filter.

A drawback of this technique is that the sky background is no longer Poisson distributed.  This is because each pixel in the diffuse background has a value less than or equal to the corresponding
pixel in the raw image.  In areas of blank sky the pixel values will be at most the mean sky value, and half the time will be lower.  The result is that in blank sky areas in the diffuse background subtracted image the mean pixel value will not be 0 but rather $\sqrt{\mu/2\pi}$, where $\mu$ is the mean number of counts of the sky background, a result which is derived in Appendix A.  Other than measuring the background flux, which can be done from  the raw image, this is not a problem.  Anywhere there is a source the background will be determined from a circular area with a large radius which will be a robust measure of the local background.  

There are other possible variants on the stacked median filter described here.
For instance, if the focus is to find absorption features in the galaxy it might be beneficial to have them appear as negative filaments in the background subtracted image.  This could be accomplished by leaving the images with the smallest filtering radii out of the stacked median filter.  However, as the purpose of this paper is to find the star forming regions above the diffuse background these possibilities are not explored further.

\subsection{Photometry\label{sec:phot}}

The analysis presented in this paper involves three different parts: photometry
of individual, resolved, star-forming regions, analysis of surface
brightness profiles, and pixel by pixel modeling of the full images.
These three analyses require different techniques. 

To identify and measure the individual star-forming regions in Holmberg IX and M81 the publicly available SExtractor \citep{SEX} code was used in conjunction
with additional routines which will be described here.

The first step is to define apertures for the star forming regions.
The UV is sensitive to star formation as the photospheres of young, massive stars are strong emitters of UV light.  Of the UVOT filters the uvm2 image would be optimal for this purpose as it is the  purest measure of
near UV light.  Both the uvw2 and uvw1 filters have extended tails in
their responses which make them subject to contamination by optical
light.  However, as shown in Figure \ref{fig:expmap} the uvw2 and uvw1
images are several times deeper than the uvm2 image.  The uvw2 image was used for source detection because it is deeper than uvm2, and because it is bluer and has less optical contamination than uvw1.

\begin{deluxetable}{ll}
\tabletypesize{\scriptsize}
\tablecaption{SExtractor parameters for M81 photometry}
\tablewidth{0pt}
\tablehead{\colhead{Parameter Name} & \colhead{Parameter Value}}
\startdata
\texttt{ANALYSIS\_THRESH} & 0.00025\\
\texttt{BACK\_TYPE} &   \texttt{AUTO}\\
\texttt{BACK\_VALUE}  &     0\\
\enddata
\tablecomments{Table \ref{tab:params} appears in its entirety in the
  online version of the \textit{Astronomical Journal}.  A portion is
  provided here for guidance regarding its form and content.}
\label{tab:params}
\end{deluxetable}

The diffuse background subtracted uvw2 image was analyzed using
SExtractor.  The SExtractor parameters used are given in Table
\ref{tab:params}, although there are a few that merit an additional
explanation.  Because the background has already been subtracted
as described above, background subtraction is turned off by setting
\texttt{BACK\_VALUE} to 0.  Although filtering images can often
improve source detections in SExtractor, in this particular case the
best results were achieved with the filtering turned off.  This is
because in many parts of the image the field is crowded.  Generally 
``Mexican hat'' filtering is useful for crowded stellar fields.
However, the star forming regions are irregularly shaped so this type
of filtering was detrimental to source detection.  Finally, the check
images were also output as they are used later in the analysis.

There are four types of magnitudes that are calculated by
SExtractor.  Because we are interested in the colors of individual
star forming regions we use the isophotal magnitudes
(\texttt{MAG\_ISO}), using the same apertures defined in uvw2 for all
images.  SExtractor will output the apertures it calculates as the
``segment map'' check image.  Each aperture is considered a segment.
In the segment map each pixel has a value of the segment number of the
corresponding aperture, or 0 if it is not in an aperture.

However there is a known shortcoming in SExtractor where clearly disjoint
objects can be grouped into the same segment \citep[e.g.,][]{Grazian06}.  In some cases an individual segment can have several components spread across
the image.  In the case of M81 these can be individual star forming
regions which are physically tens of kpc apart.  There are two schools
of thought in addressing this problem.  The first is that one of the
multiple components is usually much larger and brighter than the
others such that the effects of the smaller components are assumed to
have a minimal effect.  The second is that the segment map can be
inspected by eye and offending segments amended by hand.

The first approach seems unsatisfactory, especially since we are
attempting to determine the number and properties of the smallest star
forming regions.  On the other hand the second approach is very
tedious and fraught with the possibility of errors.  An
automated method for correcting the segment maps using Markov chains
is described in Appendix B.  The SExtractor generated segment map is corrected using this algorithm so that disjoint segments are broken up into multiple segments.

Unfortunately SExtractor does not currently have the functionality to
accept a user provided segment map.  To solve this problem an IDL code
was developed to implement the arithmetic described in Section 9 of
the SExtractor User's Manual\footnote{Version 2.5,
  \texttt{http://terapix.iap.fr/IMG/pdf/sextractor.pdf}} to determine
the positions, aperture ellipse parameters, isophotal fluxes and errors, and
background fluxes.  This also allows for a user supplied background
image.  The inputs for this code are the input image to be
photometered, the background image (in this case the diffuse background
image), the exposure map (to determine the correct exposure time as a
function of image position), the corrected segment map, and filter
file generated by SExtractor.  In this case, as the image detection
algorithm in the original SExtractor run has filtering turned off (see
Table \ref{tab:params}), the filtered image is identical to the input
image.  The filtered image is still important as it is used in the SExtractor arithmetic.  For the SDSS analysis an additional image is passed to the
code which provides the count errors as a function of the pixel.  The
count errors are calculated using the instrument counts from the
\texttt{fpC} files and the gain and noise from the sky and dark
current found in the \texttt{tsField} files.

Also included in the photometry process is an explicit correction for coincidence loss for the UVOT images.  For each source the total number of counts in the aperture and the background counts in the aperture are used to correct the count rates following the prescription in \citet{UVOTcal}.  Coincidence loss becomes significant at the 1\% level for point sources at 18.6, 18.0, and 18.5 AB magnitudes in the uvw2, uvm2, and uvw1 filters respectively.

For the SDSS data the fluxes that are output from the SExtractor
mimicking code are already in physical units given the preprocessing of
the images described above and AB magnitudes are computed directly.  For the UVOT data the count rates are converted to magnitudes using the zero points of \citet{Breeveld10b} which supersedes \citet{UVOTcal} and includes AB zeropoints.  The photometry for well detected sources is given in Table \ref{tab:photo}.

\begin{deluxetable*}{lcccccc}
\tabletypesize{\scriptsize}
\tablecaption{Source photometry and fitted parameters}
\tablewidth{0pt}
\setlength{\tabcolsep}{0.05in} 
\tablehead{\colhead{Galaxy} & \colhead{RA} & \colhead{dec} & 
    \colhead{uvw2} & \colhead{uvm2} & \colhead{uvw1} \\
 & & & \colhead{$\pm 0.03$} & \colhead{$\pm 0.03$} & \colhead{$\pm 0.03$} & \\
 & & & \colhead{0.65} & \colhead{0.69} & \colhead{0.55}}
\startdata
M81 & 149.0648 & 68.9275 & $21.39 \pm 0.03$ & $21.59 \pm 0.06$ & $21.33 \pm 0.03$ \\
M81 & 149.0598 & 68.9296 & $21.81 \pm 0.04$ & $21.99 \pm 0.07$ & $21.99 \pm 0.04$ \\
M81 & 149.0936 & 68.9309 & $22.09 \pm 0.04$ & $22.28 \pm 0.09$ & $22.47 \pm 0.05$ \\
M81 & 149.1369 & 68.9323 & $20.56 \pm 0.02$ & $20.62 \pm 0.04$ & $20.81 \pm 0.02$ \\
M81 & 149.1103 & 68.9368 & $20.93 \pm 0.02$ & $20.70 \pm 0.04$ & $20.93 \pm 0.02$ \\
\enddata
\tablecomments{See the journal article for this table.  The full data set can be retrieved from the online journal, or by emailing \texttt{hoversten@swift.psu.edu}.}
\label{tab:photo}
\end{deluxetable*}

Figure \ref{fig:expmap} shows that the exposure time in the UVOT images is highly variable across the field.  Additionally the diffuse background of the galaxy can vary on small scales as well.  As such the limiting magnitude of each filter varies across the field.  The 5$\sigma$ limiting magnitudes in the uvw2 filter are roughly 24.9 magnitudes for the parts of M81 with the least exposure time, and 25.7 in Holmberg IX.  These values are 23.7 and 24.1 in uvm2 and 24.7 and 25.8 in uvw1.

The resulting photometry is corrected for Galactic reddening by the Milky Way.  This is done using the dust maps of \citet{SFD98} and the Milky Way extinction curve of \citet{Pei92}.  The \citet{SFD98} maps have a resolution of 6.1 arcminutes and an uncertainty of 16\%.  Over the field of M81 and Holmberg IX the maps range from $A_V=0.242$ to $0.249$.  Given the uncertainty in the dust maps and the small dispersion in dust values, a single dust value of $A_V=0.248$ is used uniformly across the entire field.  This value was chosen because it is the closest value for the most pixels.  The dust correction ranges from $A_z=0.121$ at the red end to $A_{\rm uvm2}=0.693$.  The extinction is largest in the uvm2 band rather than uvw2 which is bluer because the uvm2 filter is centered on top of the 2175 \AA\ bump.

\citet{Sandage76} observed Galactic cirrus in the field of M81.  This suggests that the Milky Way extinction in this direction could have fine structure beyond the resolution of the \citet{SFD98} maps.  We acknowledge this fact, but note that our results trace features in M81 and Holmberg IX and do not appear to be affected by uncertainties in the Milky Way extinction.

For the surface brightness profile and pixel by pixel analyses the UVOT count rate images are flux calibrated using the updated photometric zero points in \citet{Breeveld10b}.

\section{Models\label{sec:models}}

Model galaxy spectra were calculated using the publicly available
PEGASE.2 spectral synthesis code \citep{PEGASE}.  Models were
calculated for ages from 1 Myr to 13 Gyr.  Twenty-five smoothly
varying star formation histories (SFH) generated by analytic formulae are considered.  The SFHs include an instantaneous burst model, 19 exponentially decaying models with time constants from 1.1 to 35 Gyr, a constant SFR, and four increasing SFHs that are proportional to $1-\exp^{-t/\tau}$, where
$\tau$ is the time constant.

Discontinuities in the SFH can mimic the effects of varying the age, stellar initial mass function (IMF), and metallicity.  Detailed modeling and reconstruction of SFHs in integrated stellar populations
is difficult, if not impossible, in many cases.  This is particularly true here
where the only constraints are eight broadband photometry points.
However this problem is mitigated for two of our techniques: our photometry of individual star forming regions, which are presumably simple stellar populations (SSP; marked by a SFH with an instantaneous burst at t=0 and no star formation after that), and for our pixel by pixel fitting, which minimize SFH variations by covering only a small amount of physical space.

The effects are more serious for the surface brightness profile analysis discussed in Section \ref{sec:surfb}.  Since the surface brightness measurements are averages over substantial fractions of the galaxy a simple SFH model no longer is good assumption.  However allowing too many parameters in the SFH modeling will quickly devolve into a degenerate problem, particularly considering that we have only 8 broadband photometry measurements as constraints.  For this reason the SFH for the surface brightness profile modeling is assumed to be either constant or exponentially decreasing with a burst of star formation of varying strength occurring at an age of 12 Gyr.  This is discussed in more detail in Section \ref{sec:surfb}.

The metallicity of the stars is assumed to be constant with respect to
time and are calculated for $Z = 0.005$, 0.010, 0.015, 0.020, 0.025,
0.035, and 0.050.  Galactic winds, galactic infall, and dust
extinction are turned off.  A wide range of dust extinction models are explored later, as there is more flexibility for doing so outside of PEGASE.  The stellar initial mass function is parameterized as 
\begin{equation}
\frac{{\rm d}n}{{\rm d}\log m} \propto \left \{ \begin{array}{ccc} -0.5 & \mbox{for}& 0.1 <m/{\rm M}_
\odot < 0.5\\
-1.35 & \mbox{for}& 0.5 < m/{\rm M}_\odot < 120 \end{array} \right.
\label{eq:imfbaldry}
\end{equation}
following \citet{BG03}.  This is similar to the \citet{S55} IMF, with
the difference being fewer stars with masses below $0.5\ M_\odot$.
There is a consensus among several authors that there is a break in
the IMF near $0.5\ M_\odot$ \citep{Kroupa01}.  Above this mass there
is a large scatter in IMF slope measurements but the mean is
consistent with the Salpeter IMF \citep{Scalo98}.  \citet{Hoversten08}
have recently shown evidence for evolution in the IMF slope tied to
galaxy luminosity.  However for the luminosity of M81 ($M_B \sim -22.7$;
\citet{Freedman01}, \citet{deV91}) it is still consistent with a
Salpeter slope.  There is also some evidence that the IMF varies with
surface brightness \citep{M09, Hoversten10}.

The final consideration in our modeling is internal dust extinction within the galaxies.  Dust values range
from $A_V = 0$ to 6 for four different dust models.  The first three
are the Milky Way (MW), Large Magellanic Cloud (LMC), and Small
Magellanic Cloud (SMC) extinction curves from \citet{Pei92}.  These
three extinction curves are very similar in the optical part of the
spectrum, but differ significantly in the ultraviolet.  The primary
difference is the strength of the bump in the extinction curves at
2175 \AA\ which is prominent in the MW extinction curve, weak in the
LMC, and largely absent from the SMC curve.  The fourth model is the
starburst dust model of \citet{Cal94} which is a grayer model for
actively star forming galaxies.

For each combination of dust model and $A_V$ the extinction curve is applied to the model spectrum output by PEGASE.2.  Synthetic colors are then determined from the extincted spectra.  The net result is a grid of models which are a function of age, SFH, metallicity, internal dust extinction, and dust model.

\section{Analysis\label{sec:analysis}}

The UVOT and SDSS imaging was used for three different analyses which are described in this section.  In section \ref{sec:sfreg} the physical properties of individual star forming regions of M81 and Holmberg IX are modeled.  Analysis of the surface brightness profiles of M81 are discussed in section \ref{sec:surfb}.  In section \ref{sec:pixel} the stellar populations of M81 and Holmberg IX are modeled on a pixel by pixel basis.  Throughout this section our analyses utilize the spectral synthesis modeling described in Section 3.

\subsection{Individual Star Forming Regions\label{sec:sfreg}}

\begin{figure*}
\plotone{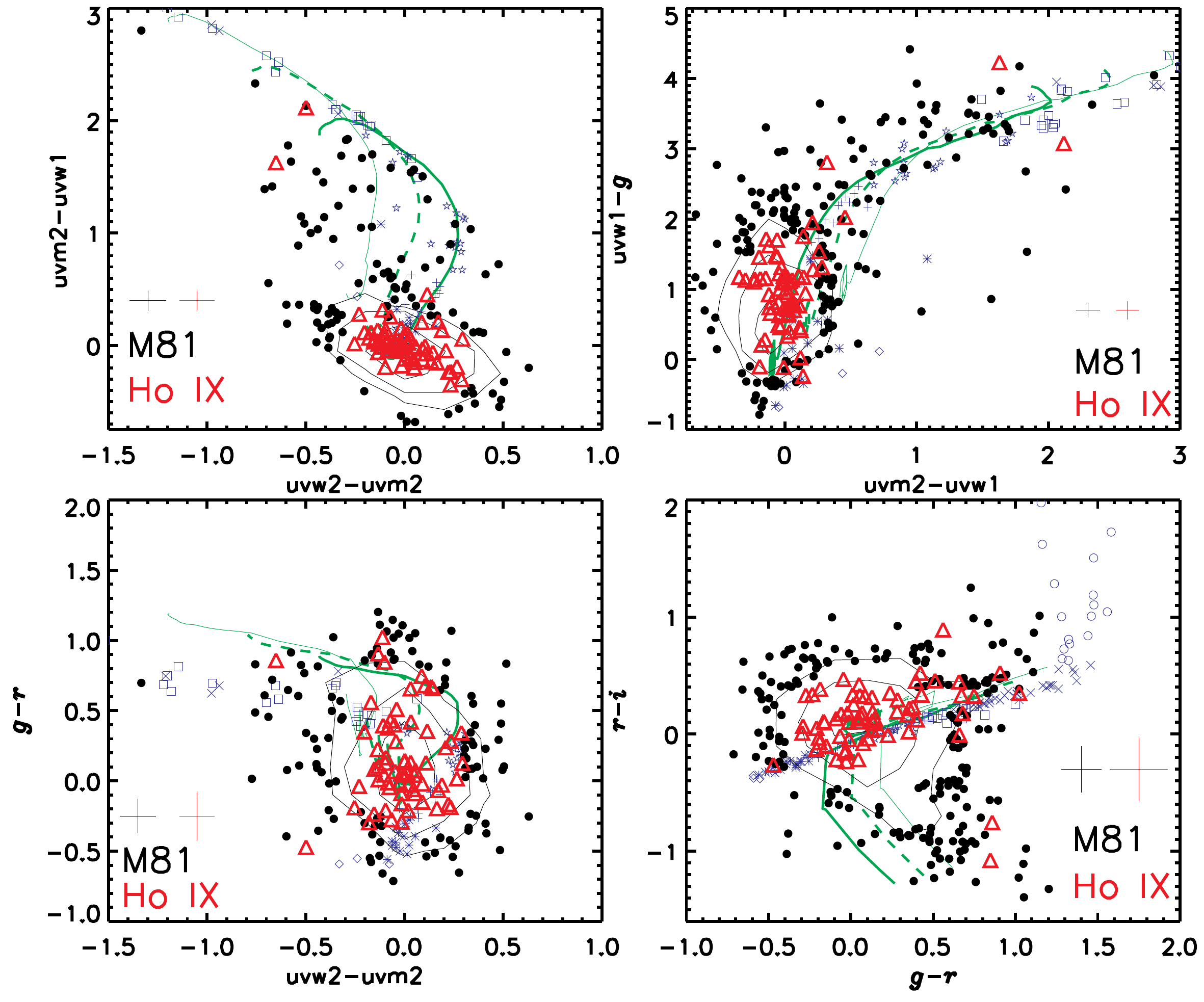}
\caption{Color-color plots of individual star forming regions in M81
  and Holmberg IX.  Photometry of M81 is shown by black logarithmic
  contours in areas of high density and black filled circles
  elsewhere.  Photometry of Holmberg IX is shown by red triangles. The median errorbars for the colors are given by the black (M81) and red (Holmberg IX) crosses.  For reference synthetic photometry of model stars from the
  \citet{Pickles} catalog is shown using a different blue symbol for spectral types O (diamonds), B (asterisks), A (plus signs), F (stars), G (squares), K (Xs), and M (circles).  Simple stellar population model tracks for solar metallicity are shown for several by the green lines assuming a Milky Way dust model for A$_V$=0.0 (thick line), A$_V$=0.5 (dashed line), A$_V$=1.0 (thin line).  The age of the stellar population evolves along the model tracks.  Top left: uvw2 $-$
  uvm2 vs. uvm2 $-$ uvw1. Top right: uvm2 $-$ uvw1 vs. uvw1 $-$ $g$.
  Bottom left: uvw2 $-$ uvm2 vs. $g-r$. Bottom right: $g-r$ vs. $r-i$.} 
\label{fig:colcol}
\end{figure*}

As described in Section 2, individual star forming regions are identified by subtracting off a model of the diffuse background light of the galaxy and using SExtractor to determine the apertures in the uvw2 band.  The diffuse background subtracted images were then photometered using the apertures from the uvw2 band.  The results of this process are shown in the color-color plots in Figure \ref{fig:colcol}.

To reduce the noise of the sample only those regions which have at least 10 ks of exposure time in both the uvw2 and uvw1 bands, and those objects which satisfy uvm2 $<23$ and $u<21.5$ magnitudes are included in the analysis.  The ground based SDSS $u$ and $z$ band imaging is significantly shallower than in other bands (the $z$ band data is of less importance for star forming regions so the magnitude cut is only placed on $u$).  The photometric measurements of the sources which satisfy these criteria are provided in Table \ref{tab:photo}.

In Figure \ref{fig:colcol} the 1260 individual star forming regions of M81 which meet these criteria are plotted as filled, black circles and logarithmic contours.  The 66 star forming regions of Holmberg IX are shown as red triangles.  For reference the blue symbols show the location of the stellar locus based on synthetic colors generated from the \citet{Pickles} database of stellar spectra.  As described in the figure caption different symbol shapes are used for the various spectral types.  Also shown in the figure are selected SSP model tracks.  The model tracks have solar metallicity and are shown for Milky Way dust models with A$_V$=0.0 (thick line), A$_V$=0.5 (dashed line), and A$_V$=1.0 (thin line).  Along the tracks the age of the SSPs evolve.

An important caveat is that the individual star forming regions were identified purely by their clumpiness above the diffuse background light.  Therefore there is likely contamination of the sample due to bright stars and globular clusters.  Furthermore the ability of SExtractor to break clumps into individual regions is limited by the resolution of the UVOT images.  For instance, given the results of \citet{Pleuss00} and \citet{Scoville01} discussed earlier the {\it HST } imaging of Holmberg IX from \citet{Sabbi08} would likely detect a larger number of smaller star forming regions.  Beyond  the resolution of the images, at some point the difference between separate star forming regions and knots of a larger complex becomes something of a philosophical question.  As such the star forming regions described in this section are defined by a best effort use of SExtractor with the resolution of UVOT which the reader is advised to keep in mind.

The top left panel of Figure \ref{fig:colcol} shows the uvw2$-$uvm2 color vs. the uvm2$-$uvw1 color.  The behavior of the stellar locus in this color space is potentially confusing.  The massive, young O type stars in this space have a color index near 0 in both uvw2$-$uvm2 and uvm2$-$uvw1.  For progressively later type stars the uvw2$-$uvm2 and uvm2$-$uvw1 colors become redder as expected.  However, for stars with spectral types later than F stars uvm2$-$uvw1 colors continue to become redder as expected, but the uvw2$-$uvm2 colors turn and become bluer.  As a result the uvw2$-$uvm2 colors of model M stars are significantly bluer than those of O stars!

This behavior is caused by an extended tail, or red leak, in the uvw2 filter.  The red leak is also present in the uvw1 filter.  This behavior is described in detail in \citet{Brown10}.  The bulk of the sensitivity of the uvw2 filter is blueward of uvm2.  However the uvm2 has a sharp cut off in sensitivity around 3000 \AA, while the sensitivity of the uvw2 filter falls off more slowly.  While the throughput of the uvw2 redward of uvm2 is low it is not zero, so objects which are bright and red can still be detected in uvw2.  As a result, the effective wavelengths of the uvw2 and uvw1 filters are particularly sensitive to the spectral shape of the object being observed.  The uvw2 filter can be either redder or bluer than uvm2, depending on the source, which explains the behavior of the stellar locus seen in Figure \ref{fig:colcol}.

Another example of red leak can be seen in the bottom right panel of Figure \ref{fig:expmap} which shows the uvm2-uvw1 color across the field of M81 and Holmberg IX.  The star forming knots have uvm2-uvw1 colors of around -0.1 to 0.2, while the color of the center of M81 is $\sim$ 1.7 to 1.8.  To get a rough estimate of the contribution from the red leak these colors can be approximately converted to stellar spectral types using Table 12 of \citet{Brown10}.  The star forming regions have roughly B and A type colors, while the galaxy center has G to K colors.  Table 12 also gives red leak corrections for the fraction of the light emitted longward of 2500 \AA\ in uvw2, and 3300 \AA\ in the uvw1 filter.  Assuming stellar spectra the red leak in the uvw2 filter is less than 0.1 magnitudes for the star forming regions, but at $\sim 1.8$ magnitudes the correction is substantial in the galaxy center.  In the uvw1 filter the red leak correction is again less than 0.1 for the star forming regions, and $\sim 0.5$ for the center of the galaxy.

These examples demonstrate that interpreting the UV flux in the uvw2 and uvw1 filters is more complicated than in the uvm2 filter.  However, all modeling in this paper uses the UVOT filter response curves.  In this way the red leak of the uvw2 and uvw1 filters is taken directly into account, rather than applying an approximate red leak correction to the flux measurements prior to the analysis.

Irrespective of the red leak issues, the principal result of Figure \ref{fig:colcol} is that the colors of the individual star forming regions in M81 and Holmberg IX are in close agreement with each other.  This suggests that the star formation in M81 and Holmberg IX is related, in agreement with theories about a past interaction with M82 and Holmberg IX's potential tidal dwarf nature.

In the top two panels describing the UV the bulk of the star forming regions have colors roughly consistent with O and B stars.  However in the bottom panels the $g-r$ colors are more consistent with A and F stars.

\begin{figure}
\epsscale{1.0}
\plotone{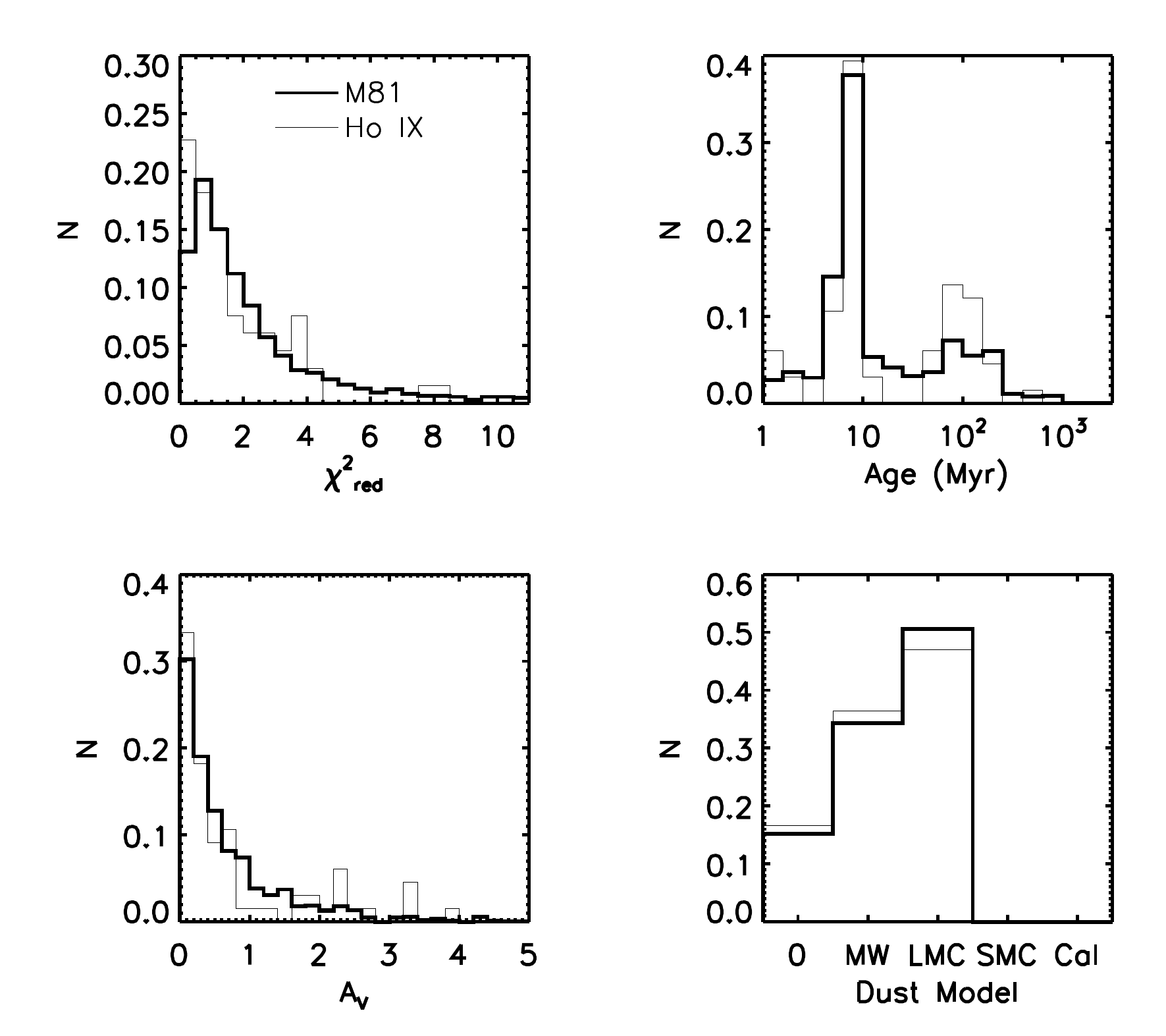}
\caption{Histograms of the best fitting parameters of individual UV
  star forming regions in M81 (thick line) and Holmberg IX (thin
  line). The derived parameters are the reduced $\chi^2$ values of the
fits (top left), the age assuming a simple stellar population (top
right), the $V$ band extinction $A_V$ (bottom left), and the best
fitting dust model (bottom right).  The dust models are those of the
Milky Way (MW), Large Magellanic Cloud (LMC), Small Magellanic Cloud
(SMC) and the starburst model of \citet{Cal94} (Cal) in order of decreasing
prominence of the 2175 \AA\ dust feature.  The ``0'' model refers to
sources where $A_V=0$ so the dust models cannot be differentiated.}
\label{fig:knotparam}
\end{figure}

While a cursory glance at Figure \ref{fig:colcol} suggests a connection between the star formation regions in M81 and Holmberg IX, this can be explored in more detail by fitting models to individual star forming regions.  To do this the 8-band photometry for each region was compared to the grid of models in Section 3 using a simple $\chi^2$ analysis.  The model parameters which minimize $\chi^2$ for each region were stored.  These best fitting parameters are plotted as histograms for the M81 and Holmberg IX regions in Figure \ref{fig:knotparam} and tabulated in Table \ref{tab:photo}.

Before analyzing the full grid of models in Section 3, physical insights were used to reduce the number of free parameters.  Because we are attempting to model individual star forming regions it is not expected that they should have complex SFHs.  Star formation is expected to occur in one short lived burst in each region.  Therefore, only models with instantaneous bursts at zero age were included in the analysis.  Secondly, the metallicity was held fixed at the solar value of $Z=0.02$.  \citet{Zaritsky94} and \cite{Kong00} both find that the metallicity of M81 is roughly Solar with a weak radial gradient.  Furthermore, this broadband analysis is insensitive to metallicity variations.  For each star forming region there are 8 photometry points and three constraints (age, $A_V$, and dust model) leaving 5 degrees of freedom.

The results of the model fitting are shown in Figure \ref{fig:knotparam}.  The thick histograms show the distribution of best fitting parameters for M81 star forming regions while the thin histograms show the results for Holmberg IX.  The histograms are normalized by the total number of regions in each sample.  The top left panel gives the reduced $\chi^2$ values showing the quality of the model fits.  At top right are the ages in Myr, at bottom left are dust extinction values, and at bottom right are the preferred extinction models.

The results shown in Figure \ref{fig:knotparam} suggest strong similarities between the star forming regions in M81 and Holmberg IX.  In both galaxies the best fitting age for the regions is predominantly between 5 and 10 Myr with a second peak around 100 Myr.  The extinction in both galaxies falls in the $0 < A_V < 1.0$ range favoring low extinction.  Neither galaxy is consistent with a weak 2175 \AA\ extinction feature as MW and LMC curves are preferred for all regions in both galaxies.

\subsection{Surface Brightness Profiles\label{sec:surfb}}

The analysis of surface brightness profiles can be used as a tool for probing radial gradients in the properties of stellar populations of a galaxy.  In this section surface brightness profiles of M81 are explored.  Given its smaller size and diffuse nature Holmberg IX does not as easily lend itself to this type of analysis as does a grand design spiral galaxy.  The spatial distribution of star forming properties in Holmberg IX is discussed further in Section \ref{sec:pixel}.

\begin{figure*}
\epsscale{1.0}
\plotone{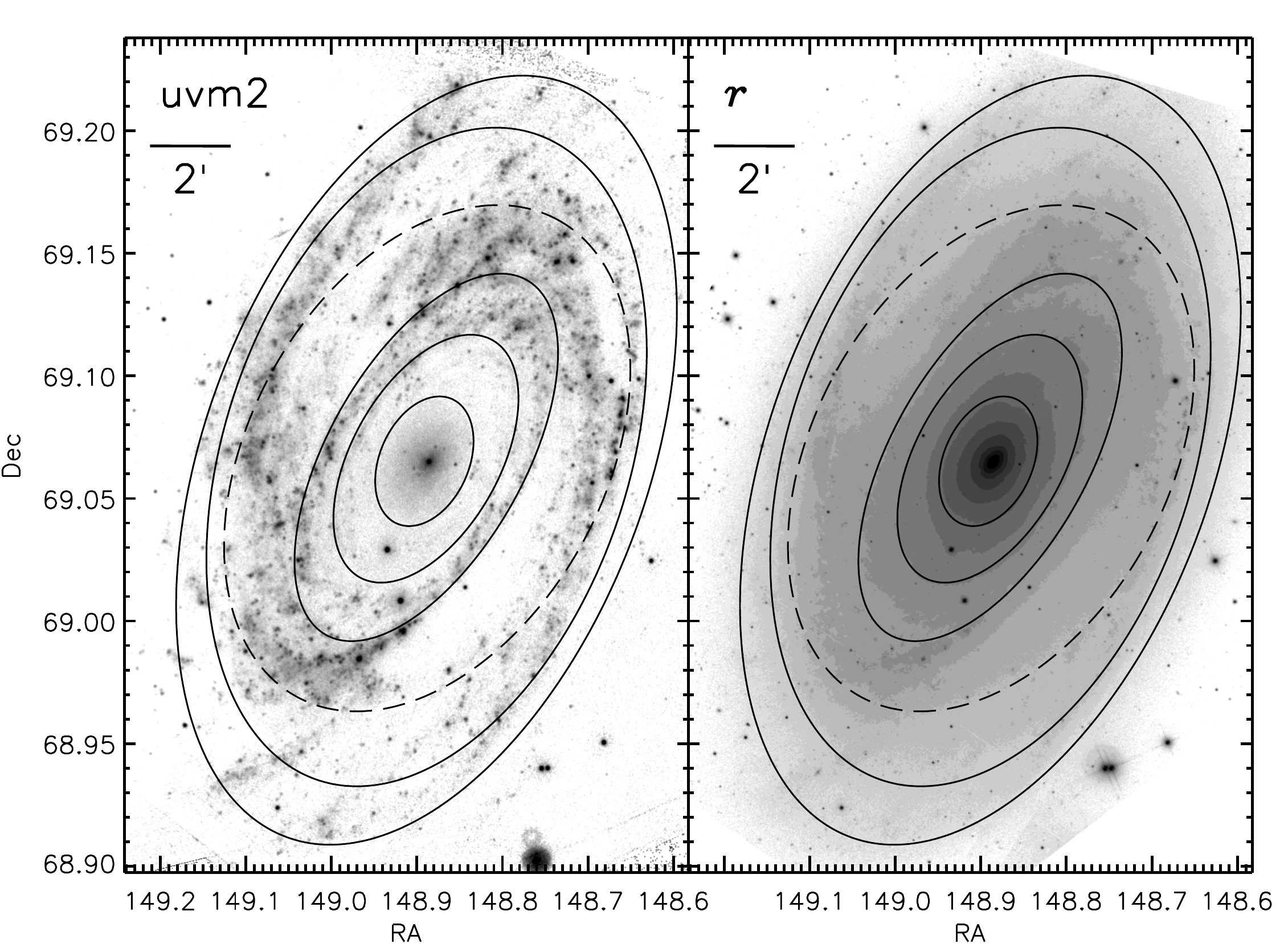}
\caption{Surface brightness contours of M81 determined from the $r$ band image with point sources removed.  The left panel shows the contours overlaid on the uvm2 image of M81, and at right the contours are shown on the $r$ band image.  The contours are identical in the two panels.  Contours are shown at intervals of 100 arcseconds in semi-major axis.  The dashed contour is at a semi-major axis of 400 arcseconds, around the point where there is an abrupt change in the elliptical isophotal parameters as shown in Figure \ref{fig:ellipse}.}
\label{fig:contours}
\end{figure*}

The surface brightness profile analysis begins with the flux corrected images described in Section \ref{sec:phot}.   Elliptical isophotes were determined using the ellipse fitting task \texttt{ELLIPSE}\footnote{\texttt{http://www.stsci.edu/resources/software\_hardware/stsdas}} within PyRAF\footnote{\texttt{http://www.stsci.edu/resources/software\_hardware/pyraf}}.  As evidenced by Figure \ref{fig:uvopt} M81 has a much smoother and well behaved radial gradient in the optical than in the UV.  The structure in UV images prevented \texttt{ELLIPSE} from calculating satisfactory isophotal profiles in the UVOT images.  In particular the ring of star formation around the nucleus seems to have caused confusion for \texttt{ELLIPSE} since the initially falling surface brightness abruptly increases.

\begin{figure}
\epsscale{1.0}
\plotone{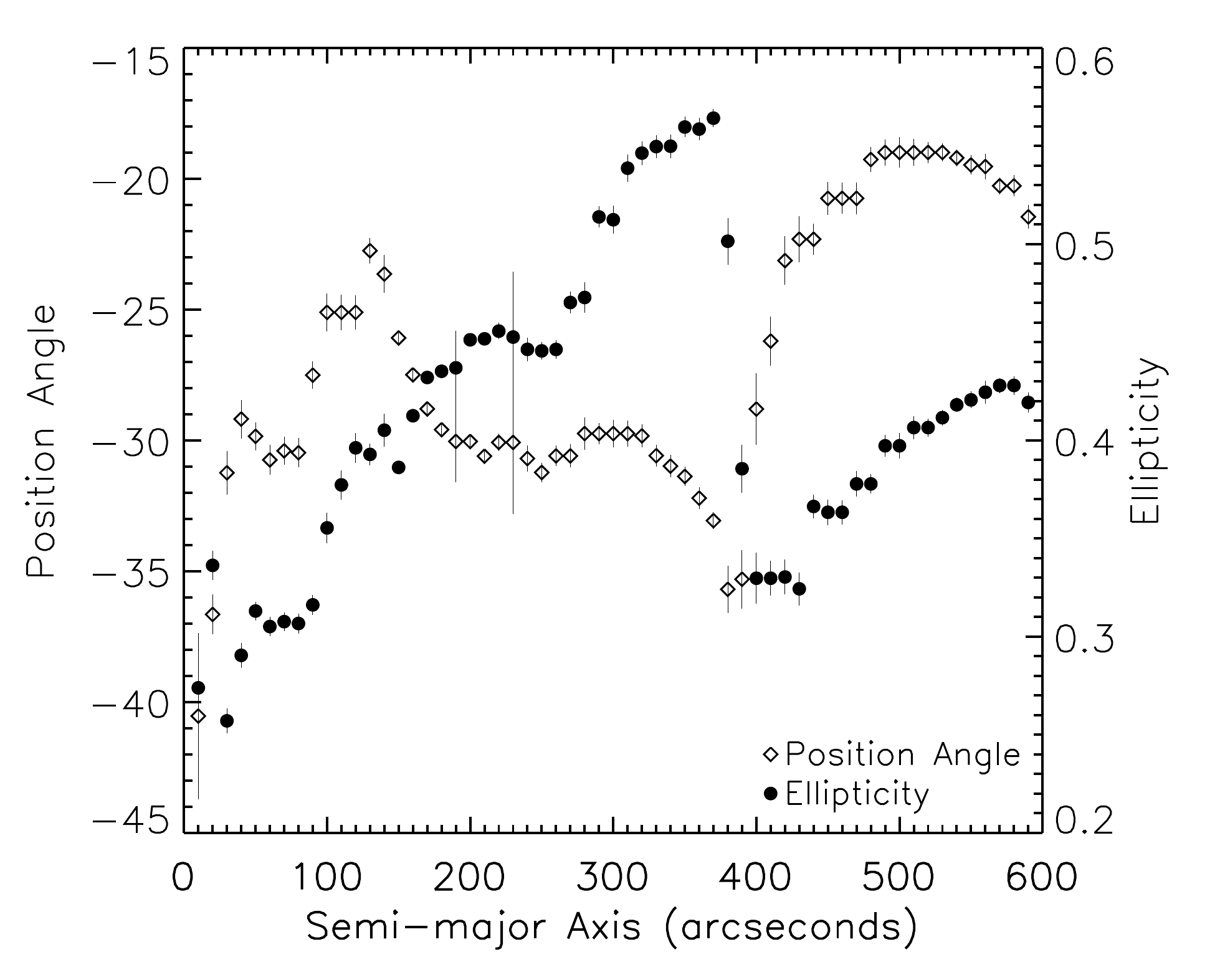}
\caption{Elliptical isophotal parameters as a function of semi-major axis as determined in the $r$ band for M81.  Diamonds indicate the position angle measured counter clockwise from north with the scaling given on the left axis.  Filled circles show the ellipticity with the scaling on the right axis.}
\label{fig:ellipse}
\end{figure}

To solve this problem, and to facilitate comparison between the optical and UV surface brightness profiles the isophotal profiles were determined using the $r$ band imaging.  These ellipses were then used to measure the surface brightness in the remaining 7 filters.  Isophotal contours are overlaid at 100 arcsecond intervals on the uvm2 and $r$ band images in Figure \ref{fig:contours}.  The measured ellipse parameters are shown in Figure \ref{fig:ellipse}.  \citet{Elmegreen95} give isophotal parameters for several optical and near infrared filters within the inner 100 arcseconds of M81.  Our isophotal ellipses within the first 100 arcseconds are in good agreement with their results at similar wavelengths.  The position angle data shows evidence of isophotal twisting, particularly in an abrupt change at 400 arcseconds.  The isophotal twisting indicates that M81 is not axisymmetric.  The ellipticity increases steadily from the middle to 400 arcseconds where it abruptly drops and then continues to increase.  The 400 arcsecond contour is denoted by the dashed contour in Figure \ref{fig:contours}.

\begin{figure}
\epsscale{1.0}
\plotone{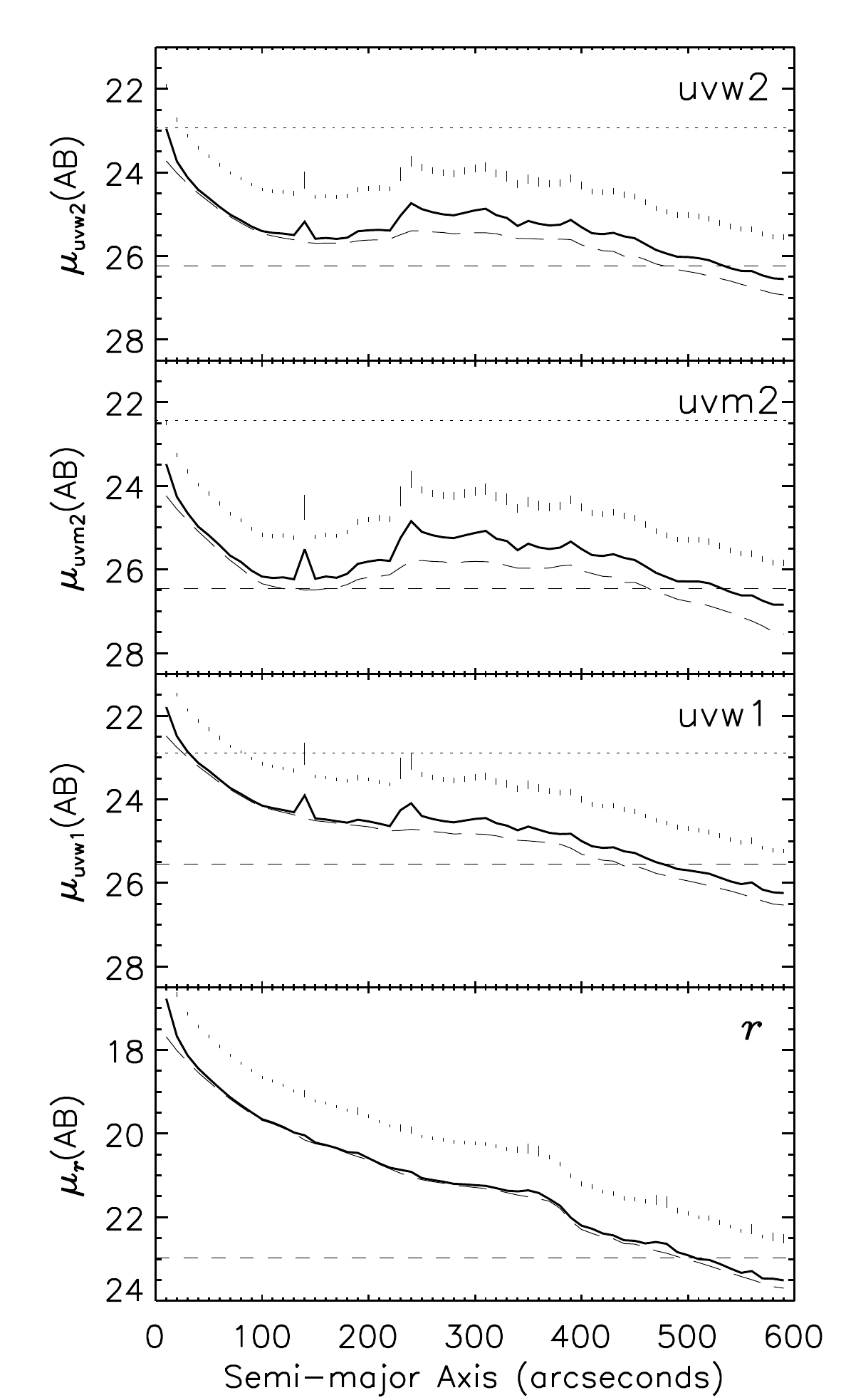}
\caption{Radial surface brightness profiles of M81 for (from top to
  bottom) the uvw2, uvm2, uvw1, and $r$ filters.  Elliptical
  isophotes were determined in the $r$ band and then used to measure
  the surface brightness in all filters.  The surface brightness,
  $\mu$ is measured in AB magnitudes per square arcsecond and the
  semi-major axes of the ellipses are measured in arcseconds.  The
  surface brightness profiles are shown for both the normal image
  (thick solid line) and the diffuse background image (thin long-dashed
  line).  The horizontal dashed line indicates the sky brightness in
  the relevant band.  The dotted line shows the surface brightness
  above which coincidence loss becomes a concern, as discussed in
  Section 2.  The errorbars for the surface brightness of the full galaxy are shown, but have been offset 1 magnitude brighter than the surface brightness for visibility.} 
\label{fig:surfb}
\end{figure}

Figure \ref{fig:surfb} shows the surface brightness profiles (thick solid line) for the UVOT filters as well as the $r$ band from which the isophotes were determined.  In addition, the thin long dashed line shows the surface brightness profiles for the diffuse background light image.  The horizontal dashed line shows the sky surface brightness and the dotted line marks the surface brightness above which coincidence loss becomes an issue, as described in Section \ref{sec:phot}.

In the $r$ band the surface brightness smoothly declines until 360 arcseconds where it drops abruptly by a magnitude before continuing a smooth decay.  Outside of the nucleus the surface brightness of the diffuse background is nearly identical to that for the whole galaxy as discrete sources do not stand out as much above the background in the $r$ band.

The situation is different in the UVOT filters.  Inside of 100 arcseconds the decline in the surface brightness is similar to that in the $r$ band.  From 100 to 180 arcseconds the surface brightness in the uvw2 and uvm2 filters flattens out to a roughly constant value and then increases by about a magnitude outside of 180 arcseconds marking the ring of star formation which can be seen in the left panel of Figure \ref{fig:uvopt}.  This increased surface brightness in the UV persists out to around 420 arcseconds before the smooth decay resumes.  The difference between the surface brightness profiles of the full and background images outside of 180 arcseconds in the top two panels shows that individual star forming regions are contributing significantly to the surface brightness in the uvw2 and uvm2 (at some radii up to a magnitude)

The uvw1 filter presents an intermediate case where the surface brightness continues to decay slowly between 100 and 180 arcseconds, and then instead of increasing, remains flat out to 300 arcseconds before continuing a slow decay.  Also it has a small drop of 0.5 magnitudes in the surface brightness at 400 arcseconds similar to that in the $r$ band.  This is due in part to the red sensitivity of the uvw1 filter as discussed above.  The uvw1 surface brightness profile shows features of both the uvm2 and $r$ band profiles located directly above and below it in the figure.

\begin{figure}
\epsscale{1.0}
\plotone{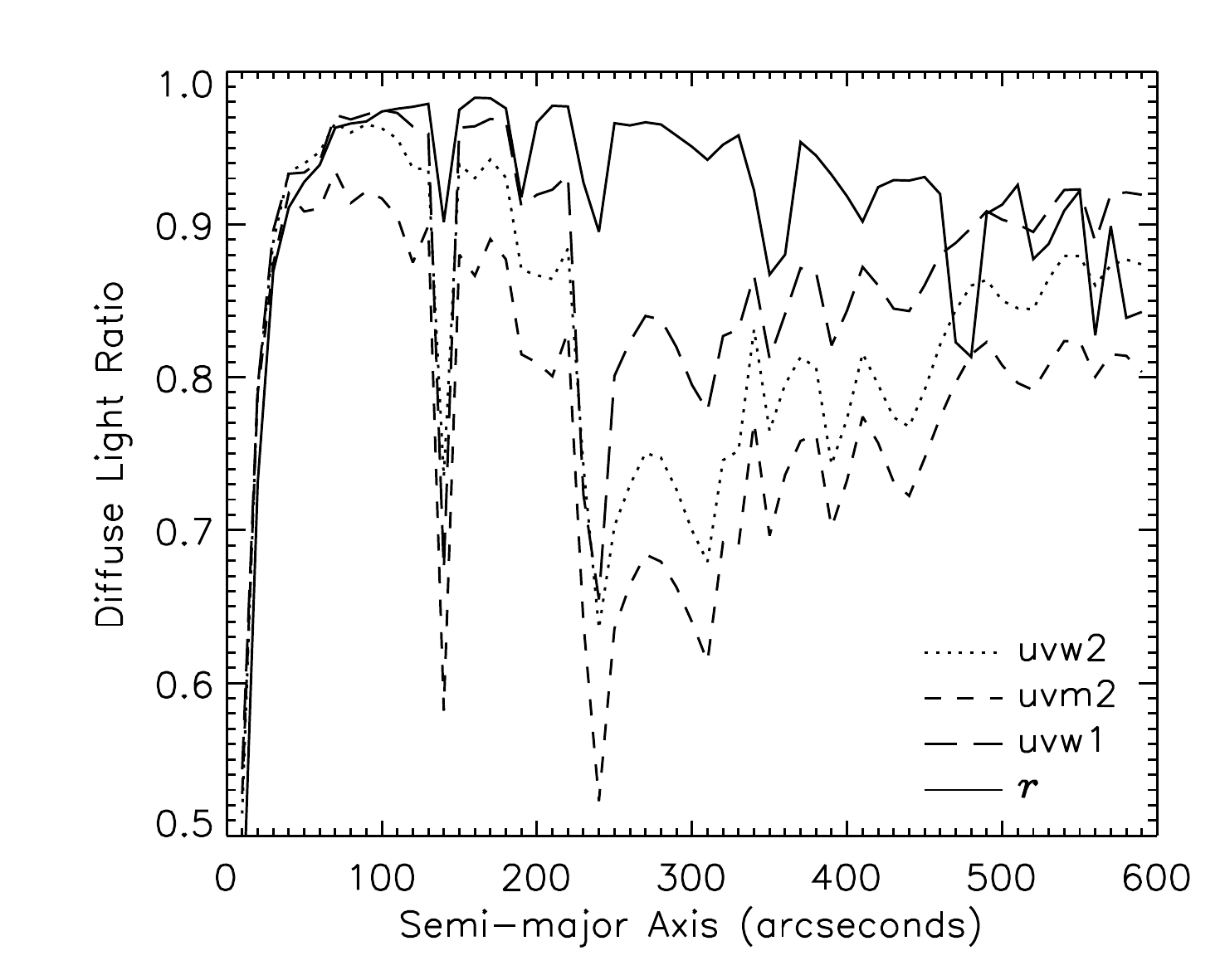}
\caption{The fraction of the total light from M81 contributed by diffuse light as a function of galactocentric radius in the uvw2 (long dash), uvm2 (dash), uvw2 (dotted), and $r$ (solid) bands.  In the UV the diffuse fraction is generally lower and reveals more structure, including a ring of star forming regions between 230 and 320 arcseconds, than in the optical.}
\label{fig:surfrat}
\end{figure}

The fraction of the total light of the galaxy contributed by diffuse light can be measured as a function of galactocentric radius using the surface brightness profiles in Figure \ref{fig:surfb}.  These diffuse light ratios are shown for several filters in Figure \ref{fig:surfrat}.  Excluding the central bulge most of the light in the galaxy comes from diffuse emission across the wavelength ranges covered in this analysis.  In the $r$ band the diffuse light fraction peaks at 98\% at around 160 arcseconds, and is followed by a slow decline in the diffuse fraction out to 600 arcseconds.  In the uvm2 band the diffuse fraction is generally lower than the $r$ band, peaking at 93\% at around 60 arcseconds, and ending up at 81\% at 600 arcseconds.  However the aforementioned ring of star formation stands out as the uvm2 diffuse fraction decreases to 52\% between around 230 to 320 arcseconds, and generally shows more structure than in the $r$ band.  The sharp dip in the diffuse fraction at 130 arcseconds is due to the foreground star GSC 04383-00565.  The diffuse light ratio in the uvw2 and uvw1 bands are qualitatively similar to uvm2, however it is higher across the board in both cases.  The dip in the diffuse fraction between 230 and 320 arcseconds reaches about 65\% in uvw2 and 66\% in uvw1.  Once again this can be understood in terms of the red leak in these filters, with uvm2 being a pure measure of the NUV light, while the red tails of the uvw2 and uvw1 filters lead to optical contamination in the measurement.

\begin{figure}
\epsscale{1.0}
\plotone{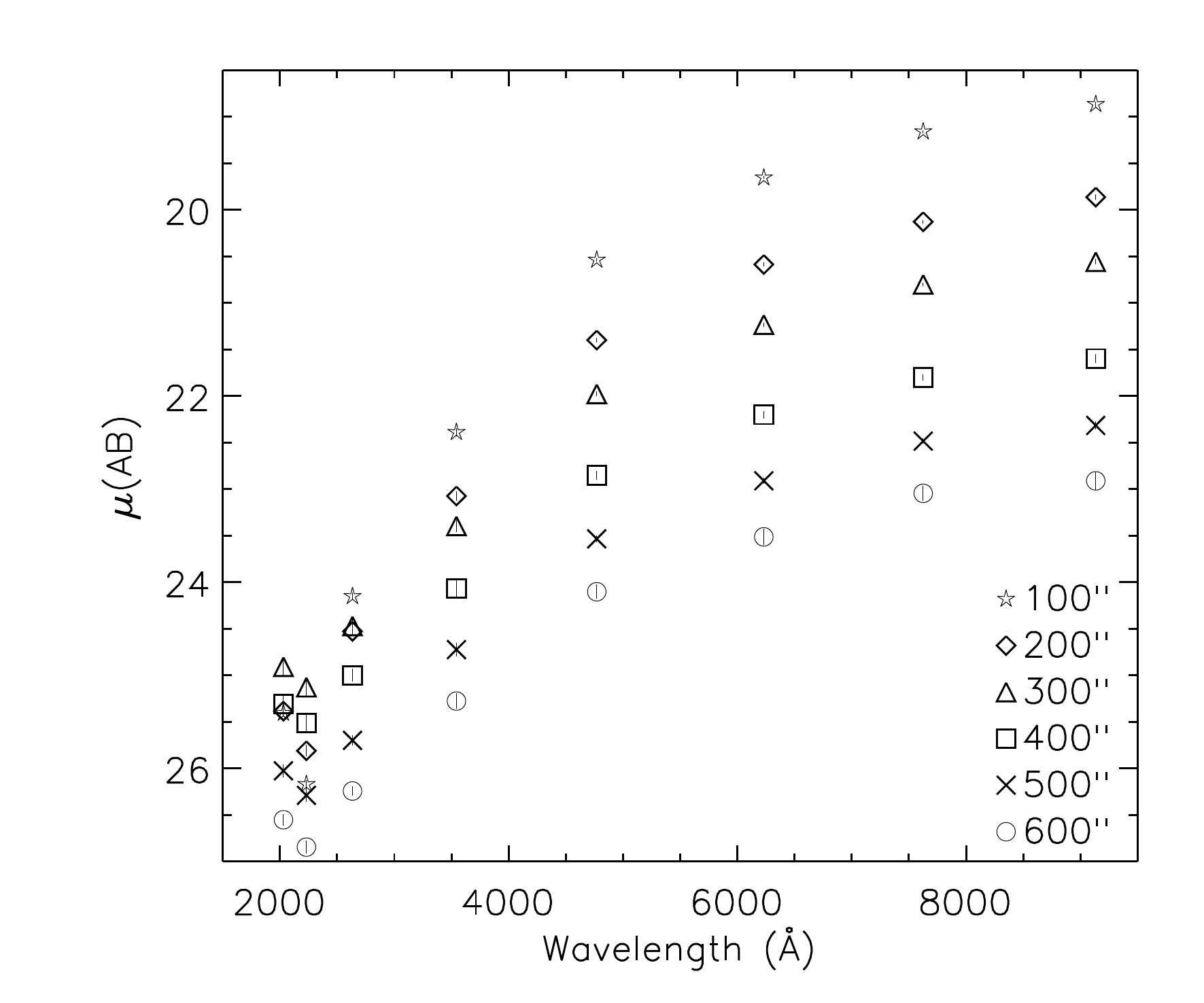}
\caption{Radial surface brightness spectral energy distributions of M81 for semi-major axes of 100 (stars), 200 (diamonds), 300 (triangles), 400 (squares), 500 (X's), and 600 (circles) arcseconds.  The surface brightness generally fades and becomes bluer with increasing radius.  Photometric errors in the surface brightness are also shown, but are generally smaller than the plotting symbols.}
\label{fig:surfc}
\end{figure}

Surface brightness profiles can be used to show how the color of the galaxy changes as a function of radius, and thus provide insight into radial changes in the stellar population.  Figure \ref{fig:surfc} shows the radial changes in the spectral energy distribution of the galaxy by using the surface brightnesses at varying semi-major axes in M81.  The general trend is that M81 gets fainter and bluer with increasing radius.

Another way to interpret the surface brightness is to fit models to the SEDs shown in Figure \ref{fig:surfc}.  This is done in much the same way as in Section 4.1, except rather than individual star forming regions the surface brightness is fit at intervals of 10 arcseconds.  The other difference is that when averaging the surface brightness over the galaxy it no longer makes sense to assume an SSP for the SFH.  A grid of SFH models was constructed, including an SSP, a constant SFR, an exponentially decreasing SFR, and constant and exponentially decreasing SFRs with star formation burst of varying strengths lasting 50 Myr at an age of 12 Gyr.

Within 180 arcseconds the best fitting models have an SSP with ages from 500 Myr to 1 Gyr and $A_V\sim 1.8$.  Outside of 180 arcseconds the best fitting models have an age of $\sim 50$ Myr after the end of a powerful burst at 12 Gyr on top of the exponentially decreasing SFR and $A_V$ from 0.5 to 1.0.  At all radii the surface brightness is best fit by a MW dust law.

Taken together the surface brightness results paint a picture of M81 in which the recent star formation is more active than it has been in the past, but that the recent star formation is only a small part of the total mass of formed stars.  At all radii the colors the profiles are too red to be described by a constant SFH at any age.  As shown by Figure \ref{fig:surfc} in the bulge of the galaxy the $r$ band light is a factor of 300 brighter than in the UV, and even out in the disk it is still a factor of 15 brighter.  The SEDs which show a red, older bulge with a younger population in the disk are in agreement with the UV image in Figure \ref{fig:uvopt}.  

\subsection{Pixel Maps\label{sec:pixel}}

The third component of the analysis of the UVOT data is pixel by pixel modeling of the images.  In this analysis each pixel of the image is treated as an individual object.  Each pixel has a calculated flux value and flux error in each of the images yielding eight total photometry points and errors for each pixel.  As in Section \ref{sec:sfreg} the photometry of each pixel is subjected to a $\chi^2$ analysis with a grid of models to determine the best fitting properties of the stellar population at that region of the galaxy.  Both M81 and Holmberg IX are modeled in this section.

Starting with the flux calibrated images described in Section 2.2, the first step in the pixel mapping is to rebin the images $4\times4$ to improve the signal to noise of the photometry in each pixel.  The $4\times4$ binning of the images that were already binned $2\times2$ onboard the spacecraft leads to an $8\times8$ binning of the images.  This corresponds to dimensions of 4 arcseconds on a side.  At the distance of M81 the physical size of the pixels is $70\times70$ parsecs.

As in Section \ref{sec:sfreg}, analysis of the individual pixels also utilized the models of Section 3.  The metallicity was again fixed to the Solar value.  The analysis was also performed with metallicity as a free parameter, but the metallicity was completely unconstrained.  However the full range of described SFHs was used for the pixel modeling.  Given that the pixels represent a 4900 pc$^2$ portion of the galaxy, including diffuse regions, dust lanes, and star forming regions, the same expectation that SSP models will apply cannot be made.  In the pixel by pixel modeling there are 8 photometry points and 4 constraints (age, SFH, extinction, and dust model) yielding 4 degrees of freedom.

In the process of the $\chi^2$ analysis, errors in the measured parameters were computed by storing all models for which the reduced $\chi^2$ of the model is within 1 of the reduced $\chi^2$ of the best fitting model.  For each parameter the range of values that meet this criteria is considered the error in that parameter.

\begin{figure*}
\epsscale{1.0}
\plotone{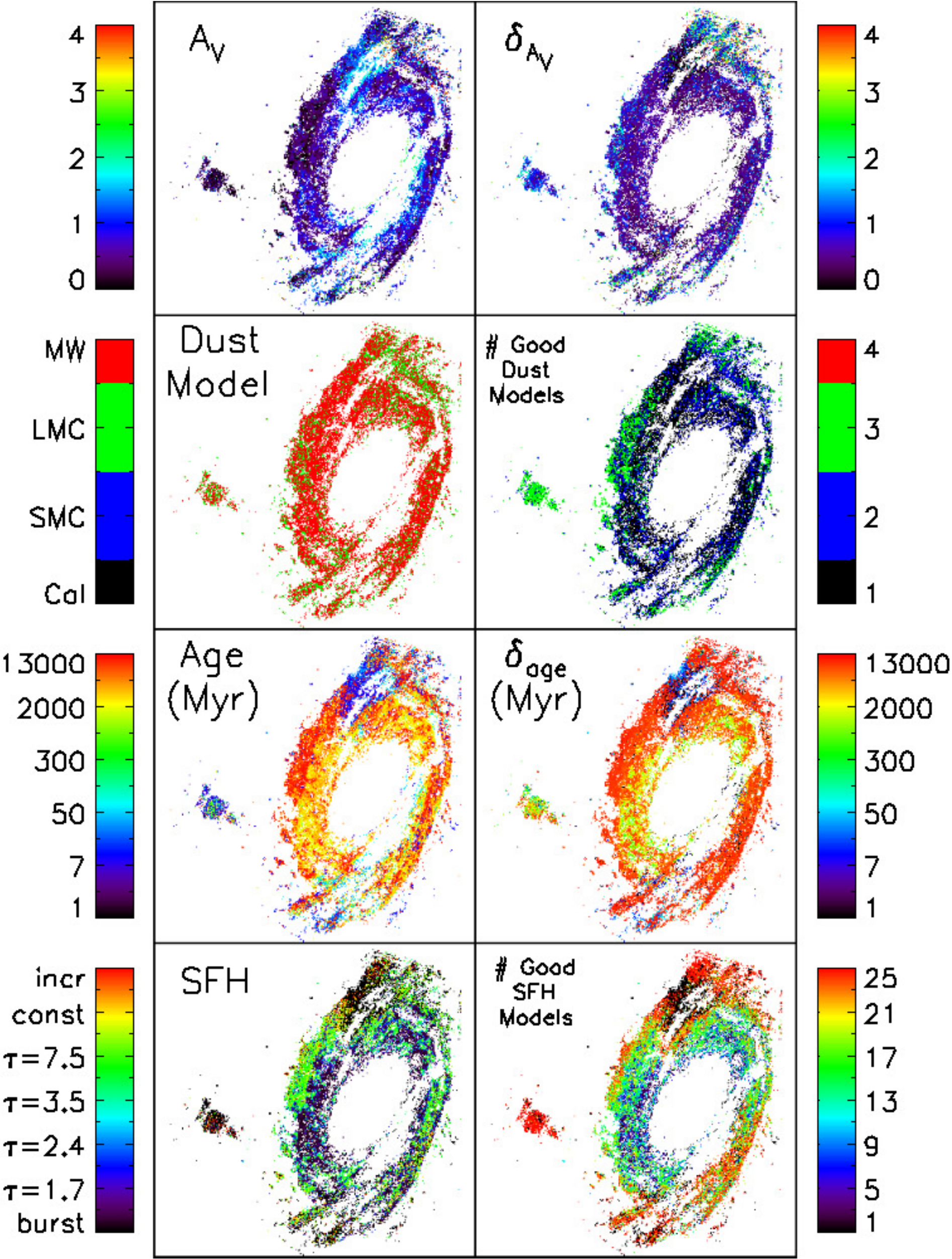}
\caption{Derived parameters of the stellar populations of M81 and
  Holmberg IX from pixel by pixel fitting. The left panels show the
  values of the measured parameters and the right panels show the
  uncertainty in those parameters as described in the text.  From top
  to bottom the measured parameters are the best fitting dust
  extinction in the $V$ band, the dust model as described in the
  caption of Figure \ref{fig:knotparam}, the luminosity weighted age
  of the stellar population, and the star formation history as
  described in the text.  Parameter values are indicated by the color
  bars at left.  The color bars at right apply to the errors in the derived parameters.  For $A_V$ and age this is self explanatory.  For dust and SFH the number of models which are successful fits to the data are shown.  For instance in the SFH panel red pixels indicate that all 25 SFH models are fit and the SFH is thus unconstrained by the fits, while black pixels indicate that only one model fits and is thus highly constrained.  White space indicates areas where the model fitting was unsuccessful, outside the analysis region, or the uvm2 surface brightness was too low for modeling.}
\label{fig:pixbypix}
\end{figure*}

Figure \ref{fig:pixbypix} shows the results of the pixel by pixel analysis.  The left hand column shows maps of the derived properties of the parameters, and the right hand column provides maps of the calculated errors of the parameters in the first column.  The values of the parameters are given by the color bars at left, while the sizes of the errors are given by the color bars on the right.  For errors in $A_V$ and age this is self-explanatory.  For the dust and SFH errors the number of good fitting models is shown.  Where the dust and SFH are tightly constrained there is only one good model.  In total there are 4 dust models and 25 SFH models.  Where the number of good models equals these numbers the dust and SFH are totally unconstrained.

White space indicates areas which have been excluded for one or more of the following three reasons.  Areas outside the analysis region are excluded, as are regions where the best reduced $\chi^2$ was large.  In addition, areas where the uvm2 surface brightness is too low are also excluded.  As can be seen in exposure maps in Figure \ref{fig:expmap}, the uvw2 and uvw1 images are considerably deeper and the uvw2 and uvw1 filters have higher throughput than the uvm2 filter.  Thus the uvm2 surface brightness cut is imposed to limit the fitting to regions where the UV emission can be fully constrained.  This is important because the uvm2 filter is particularly sensitive to the 2175 \AA\ bump, so the dust model can only be accurately constrained where the uvm2 flux is detected and not an upper limit.

The top two panels show the fitted dust parameters.  In the top row the dust extinction $A_V$ is shown.  In M81, within the spiral arms the dust extinction is mainly $A_V \sim 0.5$ with errors around 0.5 while there are a few regions where the extinction is higher.  The interarm dust lanes in M81 appear white in Figure \ref{fig:pixbypix} because of the aforementioned requirement for uvm2 surface brightness.  In Holmberg IX $A_V$ is generally less than 0.5 with larger errors around 0.8.  The pixel by pixel mapping shows that the best dust model across the bulk of both M81 and Holmberg IX is MW extinction.  
The right panel shows that even the LMC dust law is excluded across the over much of the spiral arms in M81 under the error description described above.  The dust model is less constrained in Holmberg IX, however in both galaxies the Calzetti dust models are universally ruled out.

The bottom two panels need to be taken together as there can be degeneracies between the age and SFH.  The right hand panel shows that the age is largely unconstrained over most of the disk of M81, with the exception of a young portion of the arm around 10 Myr old at the top of the image.  The inner spiral arms favor SFH with exponentially decreasing SFHs with $\tau < 2$ Gyr.  Constant SFHs are ruled out in most of these regions.  The particularly young portion of the spiral arms at top favors a SSP and is narrowly constrained.  The SFH of the outer spiral arms of M81 are poorly constrained.

The age of Holmberg IX is best fit around a few hundred Myr but the errors are on the order of a few Gyr.  The SFH of Holmberg IX is completely unconstrained.

\citet{Kong00} also performed pixel by pixel mapping of stellar population parameters of M81 (their analysis does not include Holmberg IX).  Their BATC filter system \citep{Fan96} is designed to avoid bright sky lines falling in the filter.  Their imaging included 13 filters from 3800 to 10000 \AA.  The advantage of their analysis is that they had finer resolution over optical wavelengths.  However, they did not have any UV information to better constrain the recent star formation. They also only considered SSPs.  

In their age map they get a clear radial gradient in age with the oldest populations in the center.  They also find that the spiral arms are younger than the surrounding interarm areas.  The primary differences between the \citet{Kong00} analysis and the one presented here are the assumption of SSPs and the lack of UV data.  Also, we are not able to measure the interarm regions due to the depth of the uvm2 imaging.

\section{Discussion\label{sec:discuss}}

The results of the analysis of the new UVOT imaging is in good agreement with earlier investigations.  Of primary interest is the suggested interaction of M81 and M82 220 Myr ago which may have led to the creation of Holmberg IX.  The model fits of individual star forming regions in Figure \ref{fig:knotparam} bear this out.  A visible drop off in the age histograms for both galaxies occurs at $\sim200$ Myr.  This plot also shows that the SFHs of the two galaxies are strikingly similar.

\begin{figure}
\epsscale{1.0}
\plotone{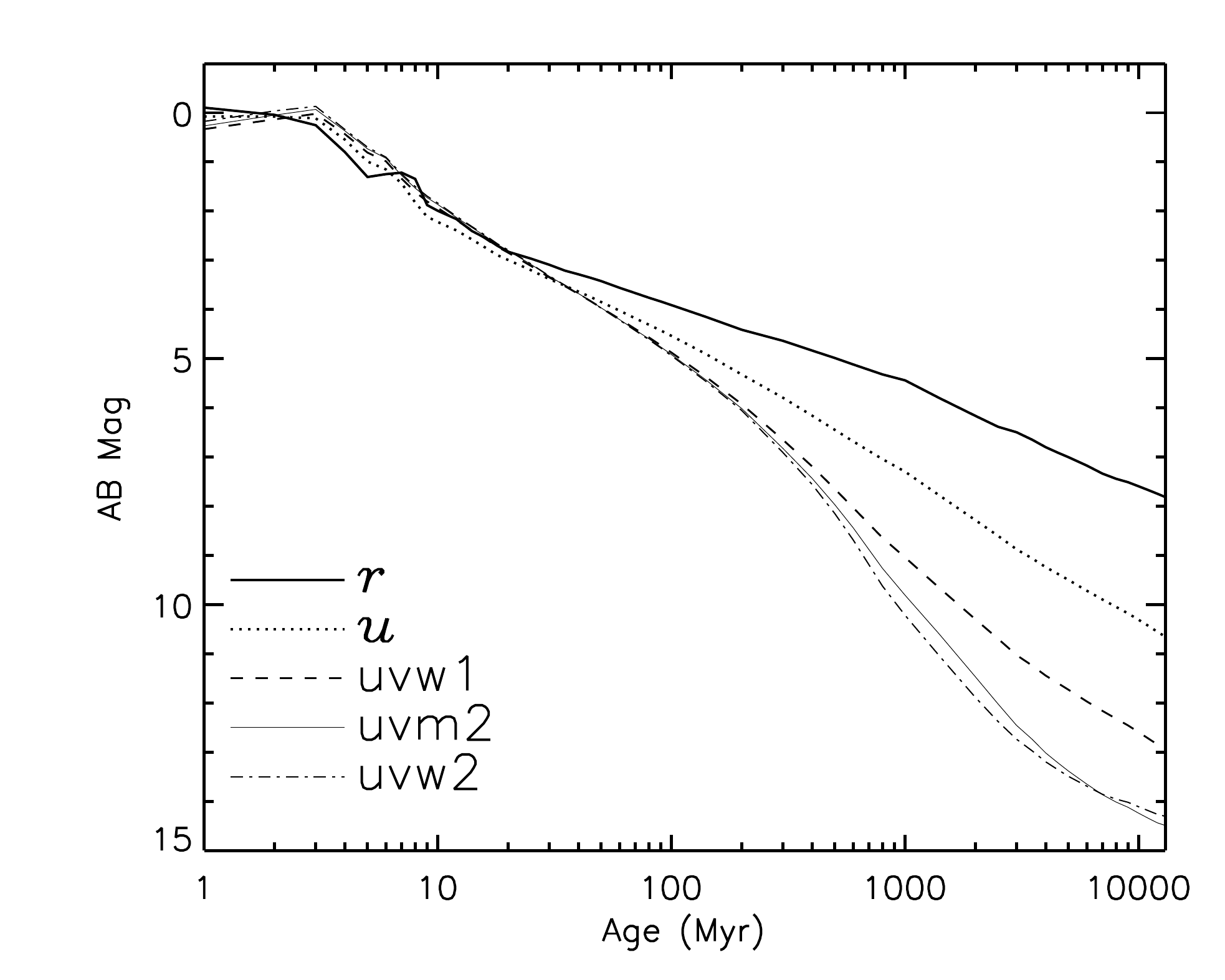}
\caption{Luminosity evolution of a simple stellar population as a function of time in five of the filters used in this paper.  The absolute magnitude is arbitrary, but the relative magnitudes of the filters is accurate.  Time is shown logarithmically in millions of years.  The filters shown are SDSS $r$ (thick line) and $u$ (dotted line) and UVOT uvw1 (dashed line), uvm2 (thin line), and uvw2 (dash-dot line).  After 20 Myr the UV luminosities drop off more quickly than in the optical.}
\label{fig:col_evolve}
\end{figure}

However, care must be taken in interpreting the age plot in Figure \ref{fig:knotparam} as there are degeneracies between the SFH and population age.  The star forming regions are identified in the uvw2 images.  Figure \ref{fig:col_evolve} shows luminosity evolution of a simple stellar population from the models described in Section 3 over time for several of the filters used in this paper.  The figure shows that in the first 10 Myr a SSP fades by 2 magnitudes from the maximum initial luminosity, and by 200 Myr it has faded by 7 magnitudes.  This heavily biases the age plot in Figure \ref{fig:knotparam} toward younger ages, which is evidenced by the largest peak at 5-10 Myr.  Nonetheless, there is still a clear drop off at $\sim200$ Myr at the expected time of the M81-M82 interaction.

This issue can be addressed to some extent by using the parameter fits of the individual star forming regions to crudely reconstruct the SFHs in the two galaxies.  A recent example of how to do this can be found in \citet{Kang09} who did so using {\it GALEX} detected UV star forming regions in M33.  To summarize, as shown in Figure \ref{fig:knotparam} there is a best fitting age, extinction, and dust model.  Using the distance to M81 and the best fitting extinction and dust model the uvm2 luminosity of each star forming region can be determined.  The simple stellar population models from PEGASE shown in Figure \ref{fig:col_evolve} can be used with the fitted age to convert the observed luminosity to a mass for the star forming regions.  One of the limitations of this method is that the mass is highly sensitive to the extinction.  For instance a $V$ band extinction of 3 magnitudes becomes almost 8 in uvm2 for a MW dust model.  The objects that were best fit by the highest $A_V$ values yield unphysical masses which skews the SFRs.  In order to combat this problem, for this experiment the analysis of Section 4.1 was repeated with a maximum allowed $A_V$ of 3.  This left the bulk of the star forming regions unaffected, and rather than lead to a pile up of sources with $A_V=3$ it forced them to choose lower extinction values with a different dust model and age.

\begin{figure}
\epsscale{1.0}
\plotone{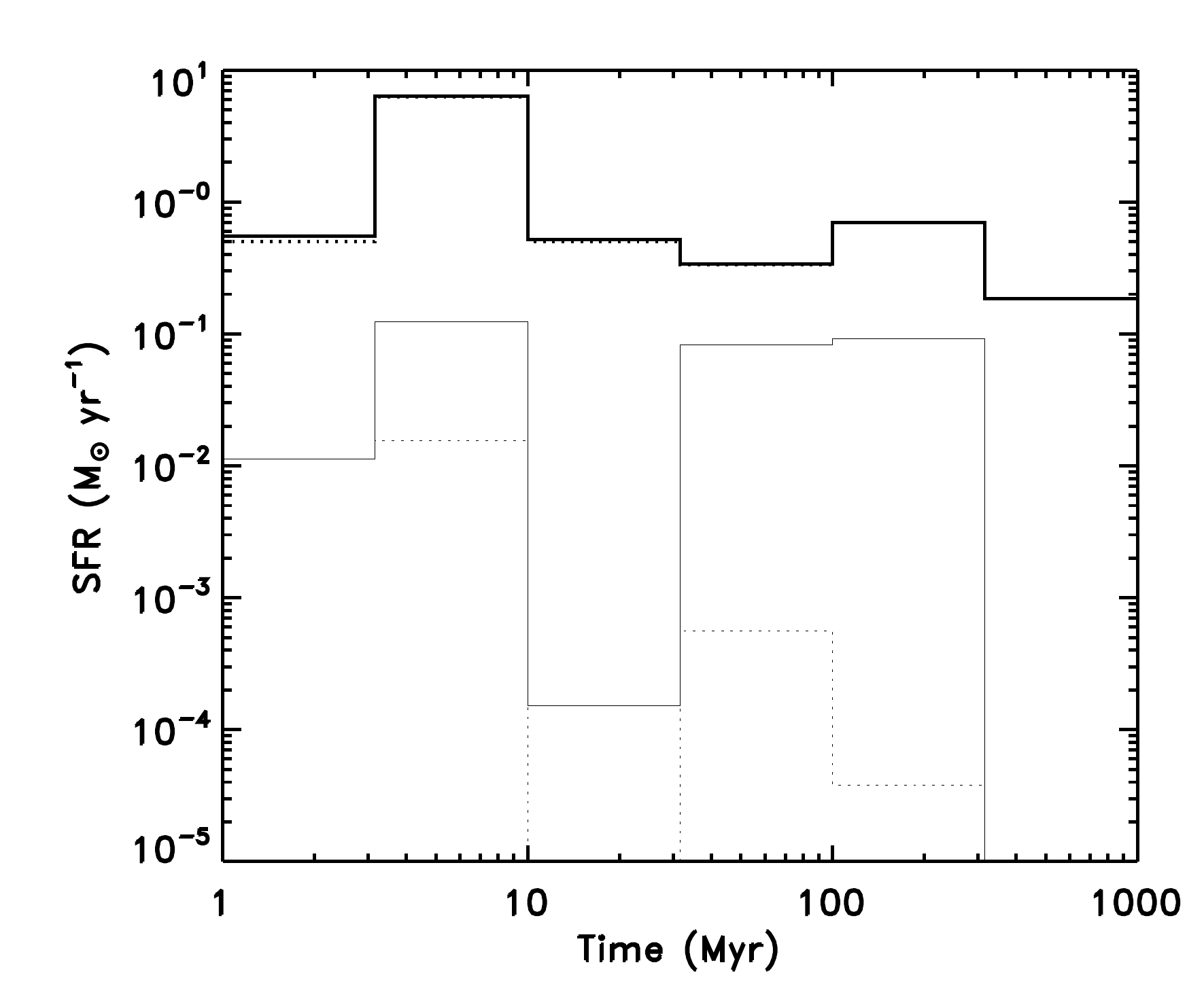}
\caption{An estimate of the recent star formation histories of M81 (thick lines) and Holmberg IX (thin lines).  The plot shows the average star formation rate in solar masses per year over logarithmic bins in time.  The solid lines show the SFR as determined from all detected UV knots.  The dotted lines show the SFR calculated from UV knots with estimated masses larger than $10^4\ M_\odot$ which is roughly luminosity limited across the time bins probed.  Both galaxies show evidence for a rise in SFR between 3 and 10 Myr ago, while Holmberg IX shows a burst of star formation lasting between 30 and 300 Myr ago with no star formation prior to that.}
\label{fig:sfr}
\end{figure}

The resulting SFHs are shown in Figure \ref{fig:sfr} by thick lines for M81 and thin lines for Holmberg IX.  The solid lines show the SFH using all detected UV knots.  However, as shown in Figure \ref{fig:col_evolve}, the UV luminosity of a star forming region decreases significantly with time.  As such, the lowest detectable mass for a region increases with time.  To correct for this, the dotted lines show the SFH constructed from regions with a minimum mass of $10^4 M_\odot$ which is roughly luminosity limited out to 500 Myr.

The current SFR of $0.5 M_\odot$ yr$^{-1}$ in M81 is consistent with the values in \citet{Gordon04}.  The SFR in M81 is shown to be roughly constant except for a factor of 10 increase between 3 and 10 Myr ago.  Being a smaller galaxy the SFR in Holmberg IX is generally lower than in M81.  A similar increase in the SFR between 3 and 10 Myr is also seen in Holmberg IX.  There is a large dip in the SFR between 10 and 30 Myr ago, and there appears to be no star formation prior to 300 Myr ago, in agreement with tidal dwarf hypotheses.  The SFR within the past 10 Myr is consistent with the values found in \citet{Weisz08}, but beyond 10 Myr it is several times higher.  Also, as shown by the dotted lines, the relative contribution by star forming regions with masses below $10^4 M_\odot$ is much more significant in Holmberg IX than in M81.  However, this could also be the result of much deeper exposure times in Holmberg IX and or low mass regions could be easier to pick out given the much simpler diffuse background in Holmberg IX.

While Figure \ref{fig:sfr} provides some interesting results, it should be interpreted with caution.  In particular, the derived masses are highly sensitive to the measured extinction, which in turn is degenerate with the fitted ages and extinction models.  Add to that the photometric errors, the number of sources in each bin, and the uncertainty in the distance to M81 and the error bars on the SFRs are not well understood.  Figure \ref{fig:sfr} is included to demonstrate a best effort at constructing SFHs from the UVOT observations.

The pixel by pixel mapping in Figure \ref{fig:pixbypix} shows that the stellar population of Holmberg IX is best fit by a stellar population with an age $\sim300$ Myr, although the age is poorly constrained.  

Perhaps the most unique contribution of this paper is the strong evidence for a Milky Way extinction curve in both M81 and Holmberg IX.  UVOT data is particularly well suited to constrain UV extinction models as the uvm2 filter is centered on and has a similar width to the 2175 \AA\ dust feature, while the uvw2 and uvw1 filters are centered at wavelengths adjacent on either side to the feature.

The results from the analysis of star forming regions in Figure \ref{fig:knotparam} shows that SMC and Calzetti dust are excluded for all individual regions in both galaxies.  In both galaxies most of the regions  are better fit by MW dust compared to the LMC model with a weaker 2175 \AA\ bump, with Holmberg IX showing a stronger bias to the MW curve.  Similarly in the pixel by pixel fitting in Figure \ref{fig:pixbypix} both galaxies prefer MW dust across their entire faces with only a small fraction of pixels where the LMC dust is preferred.  The dust model uncertainty plot shows that only the MW and LMC dust models are a good representation of the SEDs of the pixels over most of M81, and that Calzetti dust is excluded in Holmberg IX.  Together this shows that the dust in both Holmberg IX and M81 is likely similar to that here in the Milky Way.

It is useful to consider the results for the dust in M81 and Holmberg IX in the context of earlier investigations.  \citet{Roussel05} found that in NGC 300 the observed extinction laws of \ion{H}{2} regions are strongly correlated to the morphology of the regions.  In compact \ion{H}{2} regions they found that the dust showed a prominent 2175 \AA\ bump like that in the MW an LMC, but in diffuse \ion{H}{2} regions the 2175 \AA\ bump is weak as in the SMC and 30 Doradus light curves.  This is in contrast with the UVOT analysis here which found no evidence for SMC extinction regardless of morphology.  The reason for this difference is not immediately obvious.  One option is that M81 and NGC 300 have some underlying differences in their star forming regions.  Preliminary analysis as in Figure \ref{fig:pixbypix} has been repeated for UVOT observations of 6 other nearby face-on spiral galaxies in the Messier catalog.  Of the seven galaxies M81 is the most heavily tilted to a MW dust law, suggesting that its dust may have a particularly strong 2175 \AA\ bump.  Without H$\alpha$ morphologies for the UV selected star formation regions in M81 an exact comparison with \citet{Roussel05} is difficult to make.

\citet{Conroy10} looked at the \textit{GALEX} and SDSS broadband colors of a large sample of disk-dominated star forming galaxies as a function of galaxy inclination to disentangle the effects of extinction from that of the underlying stellar populations.  Their best fitting average extinction curve has $R_V \sim 2.0$ and 2175 \AA\ bump with a strength of 80\% that in the MW.  If confirmed this would be the first detection of a strong 2175 \AA\ bump in galaxies of low redshift.  \citet{Conroy10} suggest that the reason the bump is missed in earlier work is because only standard MW dust curves with $R_V=3.1$ and a strong bump are considered, which is a poor fit to the data.

The UVOT analysis here is similar to \citet{Roussel05} and \citet{Conroy10} in that it combines UV and ground based optical photometry to fit extinction curves.  However they, along with most other recent investigations of dust in galaxies at low redshift, use \textit{GALEX} FUV-NUV data to investigate the strength of the 2175 \AA\ bump.  While the UVOT lacks a FUV channel, it has the advantage of increased resolution in the NUV which makes it more directly sensitive to the effects of the 2175 \AA\ bump.  The results from UVOT for M81 and Holmberg IX point strongly to a significant bump in agreement with the recent findings of \citet{Conroy10}.

Looking forward, the analysis presented here can be repeated on a number of other nearby galaxies which have been observed by UVOT to determine the NUV star formation properties of the local Universe and to constrain dust extinction curves in galaxies of different types.


\acknowledgments

We acknowledge support from NASA Astrophysics Data Analysis grant,
\#NNX09AC87G.   This work is sponsored at PSU by NASA
contract NAS5-00136 and at MSSL by funding from the Science and
Technology Facilities Council (STFC).

Funding for the Sloan Digital Sky Survey (SDSS) and SDSS-II has been
provided by the Alfred P. Sloan Foundation, the Participating
Institutions, the National Science Foundation, the U.S. Department of
Energy, the National Aeronautics and Space Administration, the
Japanese Monbukagakusho, and the Max Planck Society, and the Higher
Education Funding Council for England. The SDSS Web site is
http://www.sdss.org/. 

The SDSS is managed by the Astrophysical Research Consortium (ARC) for
the Participating Institutions. The Participating Institutions are the
American Museum of Natural History, Astrophysical Institute Potsdam,
University of Basel, University of Cambridge, Case Western Reserve
University, The University of Chicago, Drexel University, Fermilab,
the Institute for Advanced Study, the Japan Participation Group, The
Johns Hopkins University, the Joint Institute for Nuclear
Astrophysics, the Kavli Institute for Particle Astrophysics and
Cosmology, the Korean Scientist Group, the Chinese Academy of Sciences
(LAMOST), Los Alamos National Laboratory, the Max-Planck-Institute for
Astronomy (MPIA), the Max-Planck-Institute for Astrophysics (MPA), New
Mexico State University, Ohio State University, University of
Pittsburgh, University of Portsmouth, Princeton University, the United
States Naval Observatory, and the University of Washington. 

PyRAF and STSDAS are products of the Space Telescope Science Institute, which is operated by AURA for NASA.

{\it Facilities:} \facility{Swift (UVOT)}, \facility{Sloan ()}

\appendix

\section{Diffuse Background Sky Statistics\label{sec:dbstats}}

For any photon counting instrument the sky background is expected to exhibit a Poisson distribution.  The Poisson distribution is given by
\begin{equation}
P(\nu) = e^{-\mu}\frac{\mu^\nu}{\nu!}
\label{eq:poisson}
\end{equation}
where $\mu$ is the mean number of events for a given amount of time and $P(\nu)$ is the probability of getting exactly $\nu$ events in the allotted amount of time.  The expectation value of $\nu$ can be calculated using the infinite sum
\begin{equation}
\langle \nu \rangle = \sum_{\nu=0}^\infty \nu e^{-\mu}\frac{\mu^\nu}{\nu!} = \mu
\label{eq:meanp}
\end{equation}
and the proof that $\langle \nu \rangle = \mu$ can be found in many introductory statistics textbooks \citep[e.g. ][]{Ross06}.

However for the diffuse background images described in Section \ref{sec:dback} the sky background is not Poisson distributed.  As described above, every pixel in the diffuse background image has as its value the minimum pixel value at that position in the stack of circle median filters.  In areas of blank sky the larger circle median filter sizes will have values equal to the median sky background as they take the median value over a large area of sky.  Therefore the maximum value the diffuse background image will have is the median sky value which for blank sky will be approximately the mean sky value.

In the unfiltered image, which is included in the stack, half of the sky pixels will be below the mean sky value.  For each successive radius of median filtering the value at a pixel will converge around the mean.  By picking the minimum value of the stack, the value of a pixel in the diffuse background image will follow the Poisson distribution if the observed value is below the mean sky background, and then equal to the mean if the observed value is greater than or equal to the observed background.

To calculate the expected value of the sky background in the diffuse background images Equation \ref{eq:meanp} is changed to 
\begin{equation}
\langle \nu \rangle = \sum_{\nu=0}^\mu \nu e^{-\mu}\frac{\mu^\nu}{\nu!} + \\
	\sum_{\nu=\mu+1}^\infty \mu e^{-\mu}\frac{\mu^\nu}{\nu!}
\label{eq:meanhp}
\end{equation}
to reflect that the distribution is Poisson up to the mean sky background when $\nu \leq \mu$, and then equal to the mean when the count rate exceeds the sky background ($\nu \geq \mu+1$).

To evaluate $\langle \nu \rangle = \mu$ the second sum of Equation \ref{eq:meanhp} can be rearranged as shown in Equation \ref{eq:rh1}.  In the first step top and bottom are multiplied by $\nu+1$ and the $\mu$ out front is moved into the exponential.  In the second step occurrences of $\nu+1$ are replaced by $\nu$ and the values of $\nu$ over which the summation is taken are increased by one.

\begin{equation}
\sum_{\nu=\mu+1}^\infty \mu e^{-\mu}\frac{\mu^\nu}{\nu!} = \\
	\sum_{\nu=\mu+1}^\infty (\nu+1) e^{-\mu}\frac{\mu^{\nu+1}}{(\nu+1)!} = \\
	\sum_{\nu=\mu+2}^\infty \nu e^{-\mu}\frac{\mu^\nu}{\nu!}
\label{eq:rh1}
\end{equation}

At this point Equation \ref{eq:meanhp} is identical to Equation \ref{eq:meanp} except the $\nu = \mu+1$ value is left out of the summation.  This can be fixed by both adding and subtracting the  $\nu = \mu+1$ to the right hand side
\begin{equation}
\langle \nu \rangle = \sum_{\nu=0}^\mu \nu e^{-\mu}\frac{\mu^\nu}{\nu!} + \\
	\sum_{\nu=\mu+1}^\infty \nu e^{-\mu}\frac{\mu^\nu}{\nu!} - \\
	 \left.  \nu e^{-\mu}\frac{\mu^\nu}{\nu!} \right \vert_{\nu=\mu+1}
\label{eq:rh2}
\end{equation}
such that now the summation is taken from $\nu=0$ to infinity.  The first two terms on the right hand side of Equation \ref{eq:rh2} combine to be identical to Equation \ref{eq:meanp} and are thus equal to $\mu$.  The third term is then evaluated for $\nu = \mu+1$ leaving
\begin{equation}
\langle \nu \rangle = \mu - e^{-\mu}\frac{\mu^{\mu+1}}{\mu!} \approx \mu - \sqrt{\frac{\mu}{2\pi}}
\label{eq:rh3}
\end{equation}
By invoking Stirling's Approximation, $k! \approx k^{k+1/2}e^{-k}\sqrt{2\pi}$, in Equation \ref{eq:rh3} the expected value of the sky background in the diffuse background image is shown to be less than the mean sky background by $\sqrt{\mu/2\pi}$.

Thus in the diffuse background subtracted images the mean sky background will be $\sqrt{\mu/2\pi}$.

\section{Markov Chain Segment Map Correction\label{sec:segmap}}

\begin{figure}
\epsscale{0.5}
\plotone{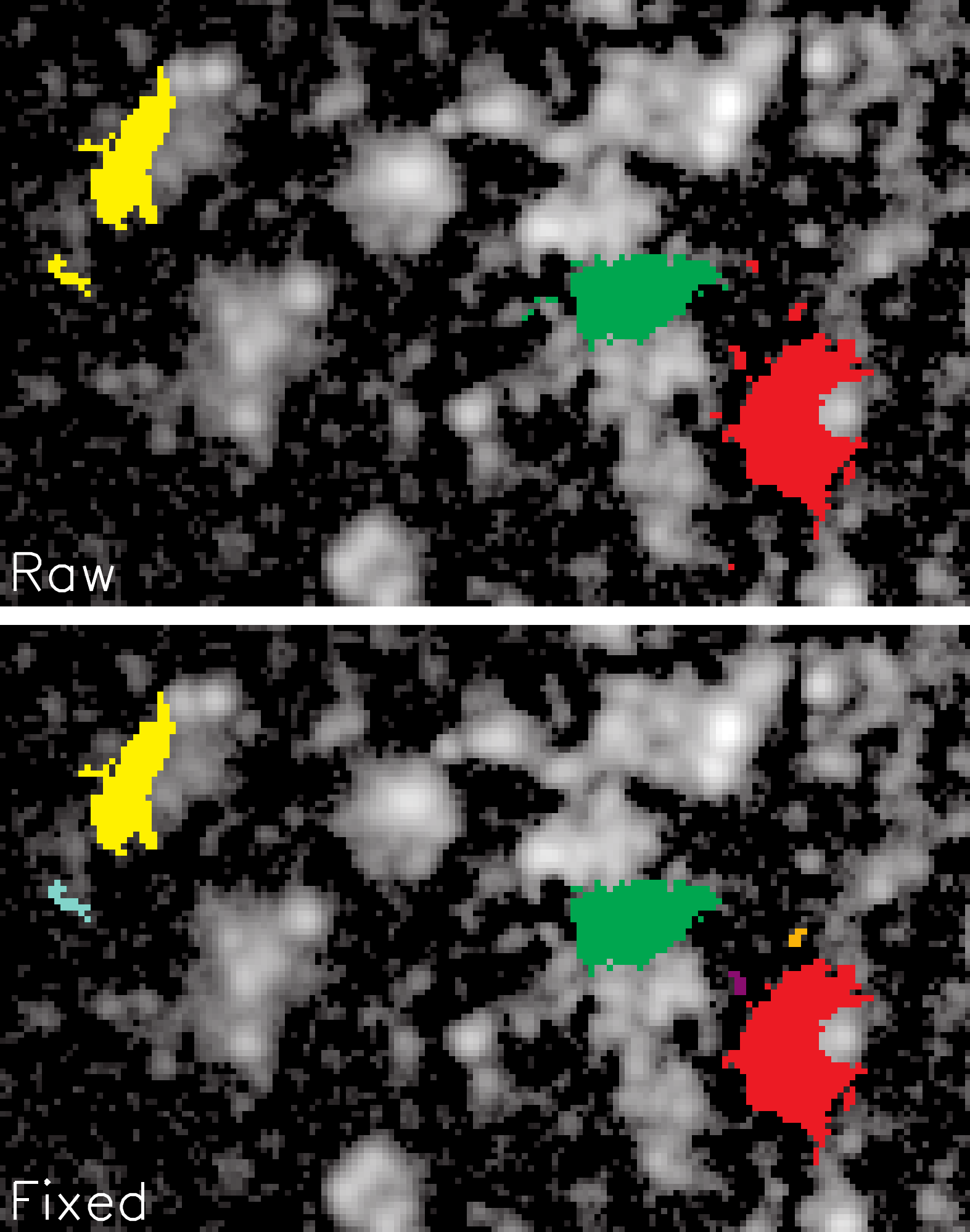}
\caption{Segment maps overplotted on diffuse background subtracted uvw2 images of a portion of the spiral arms of M81.  At top are three representative segments output by SExtractor represented by the red, green, and yellow pixels.  Pixels of the same color belong to the same segment.  At bottom are the segments after processing using the Markov chain algorithm described in Appendix B.  Three new segments have been created from the original three segments and small, detached groupings of pixels have been removed from the segments.}
\label{fig:segmap}
\end{figure}

As described in Section \ref{sec:phot} there is a known problem that can occur in SExtractor segment maps where disjoint regions, sometimes separated by large numbers of pixels, are erroneously identified as belonging to the same source.  This problem is demonstrated in Figure \ref{fig:segmap} for our M81 imaging.  In the top panel three representative segments output by SExtractor are plotted on a diffuse background subtracted uvw2 image of a portion of the spiral arms of M81.  Pixels of the same color are considered to be part of the same object by SExtractor.  All three of the segments, while dominated in these cases by main sections, contain multiple disjoint components.  Some of the disjoint components are clumps of many pixels while others are comprised of a single or a few pixels.  Each pixel in the image has a physical size of 17 parsecs so the real separation of the components is in some cases 200 parsecs or more for the three segments shown here and is worse in other parts of the image.

The two common solutions to this problem are either to adjust the segments manually or to make the assumption that most of the light will fall in the main portion of the segment and the problem can be ignored.  However the number of segments detected for this image is several thousand so a manual solution is not attractive.  While the second option may be reasonable for the red and green segments in the figure, the yellow segment should clearly be divided in two.  In this appendix we describe an automated procedure for correcting SExtractor segment maps using Markov chains.  A description of the basic equations and properties of Markov chains can be found in many introductory statistics texts, such as \citet{Ross06}.

The problem is to determine whether or not a pixel in a given segment is connected to any other pixel in the segment.  This solution proposed here uses a random walk to determine the connectivity between pixels within a segment.  The question is essentially whether a path can be drawn between two pixels in a segment by passing only through pixels within the segment.

This is implemented by analyzing the segments individually.  First the $m$ pixels within a segment are labeled 1 through $m$.  Then an $m\times m$ transition probability matrix is constructed for the segment such that
\begin{equation}
p_{ij} = P(X_1 = j | X_0 = i) = \left \{ \begin{array}{cc} 0 & {\rm if\ } i,j {\rm \ not\ adjacent}\\
1/a &   {\rm if\ } i,j {\rm \ adjacent} \end{array} \right.
\label{eq:markov}
\end{equation}
where $p_{ij}$ is the probability that the random walk will be in pixel $j$ given that in the previous step it was in pixel {i}, and $a$ is the number of pixels adjacent to pixel $i$ which are in the segment.  The value $a$ can range from 0 to 8 as we consider pixels which share a corner to be adjacent.  Because $p_{ij} \geq 0$ for all $i,j = 1,2,...,m$ and $\sum_{j=1}^m p_{ij} = 1$ and the number of pixels in the segment is finite $p_{ij}$ is by definition a Markov chain.  The probability that a walk that started in pixel $i$ will be  in pixel $j$ after $n$ steps is $p_{ij}^{(n)}$ which is the result of $n$ matrix multiplications of the matrix $p$ with itself.

We want to find the probability that a walk that starts in pixel $i$ will be in pixel $j$ after an infinite number of steps to determine whether the two pixels are connected within the segment.  It turns out that for many Markov chains $p_{ij}^{(n)}$ converges as $n \rightarrow \infty$, although it is not in general guaranteed to happen. The question then becomes how large must $n$ be to reach a suitable approximation of the probability matrix after an infinite number of transitions.

To determine when a suitable approximation has been achieved we use the variation distance
\begin{equation}
\left \Vert p^{(n)} - p^{(n-1)} \right \Vert \equiv \max_{1\leq i,j \leq m} \left \vert p_{ij}^{(n)} - p_{ij}^{(n-1)} \right \vert
\label{eq:stop}
\end{equation}
which is the maximum absolute error in any of the individual transition probabilities caused by approximating the $n$th step with the $(n-1)$th step \citep[e.g. ][]{Levin08}.  A stopping condition is imposed by requiring that the variation distance in Equation \ref{eq:stop} be less than a value $c$ which in our code is set to $10^{-8}$.

The transition probability matrix $p^{(n)}$ can then be used to determine whether all pixels in a segment are connected.  If $p_{ij}^{(n)} > 0$ for all $i$  and $j$ then all pixels in the segment belong to one object and the segment map can be left as is.  If that is not the case $p^{(n)}$ can be used to identify clumps of pixels which have a finite probability of transitions between each other and no chance of transitions to the other groups of pixels.  As required in the initial SExtractor run (see Table \ref{tab:params}) there must be at least six adjacent pixels above the threshold to be identified as a source.  Clumps of pixels with less than six members are deleted from the segment.  Clumps of more than six pixels are assigned a new segment number, breaking the original segment into two or more pieces.

The bottom panel of Figure \ref{fig:segmap} shows the results of this process.  The yellow and red segments from the top panel have been broken into two and three objects respectively and smaller groupings of unconnected pixels have been removed.  The fixed segment map is then used as input in a program which applies the SExtractor arithmetic resulting as described in Section \ref{sec:phot}.

Based on the analysis of UV observations of M81 and Holmberg IX in this paper the segment map cleaning algorithm described here worked very well.  Of the 1260 star forming regions in M81 described in Section 4.1, only 11 were created by the cleaning algorithm-- less than 1\%.  Nine sources were divided into two by the algorithm, with one being split into three sources.  Each of the 10 sources split by the algorithm was inspected by eye with all being correctly divided.  None of the 66 sources in Holmberg IX were created by the algorithm.

The one caveat to keep in mind for this technique is that the matrix multiplication computation time for an $m\times m$ matrix scales as $m^3$.  For segments with over a few thousand pixels this can become prohibitively long.  This is a problem for the bulge of M81 which SExtractor treats as a single object using the parameters of Table \ref{tab:params}.  To avoid this problem the code is set to skip segments with large numbers of pixels.

\end{document}